\documentclass[12pt]{article}
\pdfoutput=1
\usepackage{latexsym, graphicx,cite} 
\usepackage{amsmath}
\usepackage{amssymb}

 \newcommand{\arXiv}[1]{\href{http://www.arXiv.org/abs/#1}{#1}}
\usepackage[colorlinks=true, linkcolor=blue, bookmarks=true]{hyperref}

\makeatletter
\renewcommand\section{\@startsection {section}{1}{\z@}%
                                   {-3.5ex \@plus -1ex \@minus -.2ex}%nn
                                   {2.3ex \@plus.2ex}%
                                   {\normalfont\large\bfseries}}
\renewcommand\subsection{\@startsection{subsection}{2}{\z@}%
                                     {-3.25ex\@plus -1ex \@minus -.2ex}%
                                     {1.5ex \@plus .2ex}%
                                     {\normalfont\bfseries}}
\makeatother

\newcommand{\beq}{\begin{equation}}
\newcommand{\eeq}{\end{equation}}
\newcommand{\ber}{\begin{array}}
\newcommand{\eer}{\end{array}}

\newcommand{\del}{\partial}

\newcommand{\ssty}{\scriptstyle}

\newcommand{\cnst}{\mbox{const}}

\newcommand{\eps}{\varepsilon}

\newcommand{\om}{\omega}

\newcommand{\ena}{\end{eqnarray}}
\newcommand{\beqa}{\begin{eqnarray}}
\newcommand{\eeqa}{\end{eqnarray}}
\newcommand{\bea}{\begin{eqnarray}}
\newcommand{\eea}{\end{eqnarray}}

\newcommand{\be}{\begin{equation}}
\newcommand{\ee}{\end{equation}}

\begin{document}
\begin{titlepage}
\begin{flushright}
\phantom{arXiv:yymm.nnnn}
\end{flushright}
\vfill
\begin{center}
{\Large\bf Renormalization, averaging, conservation laws\\ and AdS (in)stability}    \\
\vskip 15mm
{\large Ben Craps$^a$, Oleg Evnin$^{b,a}$ and Joris Vanhoof$^a$}
\vskip 7mm
{\em $^a$ Theoretische Natuurkunde, Vrije Universiteit Brussel and\\
The International Solvay Institutes\\ Pleinlaan 2, B-1050 Brussels, Belgium}
\vskip 3mm
{\em $^b$ Department of Physics, Faculty of Science, Chulalongkorn University,\\
Thanon Phayathai, Pathumwan, Bangkok 10330, Thailand}

\vskip 3mm
{\small\noindent  {\tt Ben.Craps@vub.ac.be, oleg.evnin@gmail.com, Joris.Vanhoof@vub.ac.be}}
\vskip 10mm
\end{center}
\vfill

\begin{center}
{\bf ABSTRACT}\vspace{3mm}
\end{center}

We continue our analytic investigations of non-linear spherically symmetric perturbations around the anti-de Sitter background in gravity-scalar field systems, and focus on conservation laws restricting the (perturbatively) slow drift of energy between the different normal modes due to non-linearities. We discover two conservation laws in addition to the energy conservation previously discussed in relation to AdS instability. A similar set of three conservation laws was previously noted for a self-interacting scalar field in a non-dynamical AdS background, and we highlight the similarities of this system to the fully dynamical case of gravitational instability. The nature of these conservation laws is best understood through an appeal to averaging methods which allow one to derive an effective Lagrangian or Hamiltonian description of the slow energy transfer between the normal modes. The conservation laws in question then follow from explicit symmetries of this averaged effective theory.

\vfill

%\begin{flushleft}
%PACS 11.25.-w, 04.65.+e
%\end{flushleft}
\end{titlepage}

%%%%%%%%%%%%%%%%%%%%%%%%%%%%%%%%%%%%%%%%%%%%%%%%%%%%
%%%%%%%%%%%%%%%%%%%%%%%%%%%%%%%%%%%%%%%%%%%%%%%%%%%%
%%%%%%%%%%%%%%%%%%%%%%%%%%%%%%%%%%%%%%%%%%%%%%%%%%%%

\section{Introduction}

Non-linear instability of anti-de Sitter (AdS) space has attracted a considerable amount of attention since the pioneering observations of \cite{Bizon:2011gg}. Reasons include the inherent mathematical depth of the problem and its dual interpretation in terms of thermalization processes in quantum gauge theories (in the context of the AdS/CFT correspondence). AdS space is known to be linearly stable, but all frequencies of normal modes are integer in appropriate units, in which case non-linearities are known to induce significant (perturbatively) slow resonant transfer of energy between different normal modes, no matter how small the perturbation amplitudes are.

The ultimate fate of the energy flow between the different normal modes induced by non-linearities is a subtle matter \cite{Bizon:2011gg,Dias:2011ss,Dias:2012tq,deOliveira:2012dt,Maliborski:2013jca,Buchel:2013uba,Bizon:2013xha,Abajo-Arrastia:2014fma,Maliborski:2014rma,Balasubramanian:2014cja,Craps:2014vaa,Dimitrakopoulos:2014ada,Buchel:2014dba}, and most of the available considerations are numerical. At first, a series of numerical examples of smooth initial data that develop a turbulent cascade leading to energy transfer to very short wave length modes and black hole formation was presented in \cite{Bizon:2011gg}. It was later observed, however, that other initial profiles do not lead to collapse \cite{Maliborski:2013jca,Buchel:2013uba} and that some explicit finite deformations of AdS make it stable \cite{Dias:2012tq}. This seems to imply a complicated interplay of stable and unstable behavior described by a rich topography in the phase space. It was further suggested in \cite{Dimitrakopoulos:2014ada} that the instability domain might even shrink to a set of measure zero as the perturbation amplitude is decreased, but more analysis will be required to either confirm or rule out this possibility.

The subtleties of the AdS stability phenomena make it necessary to go beyond the inherent uncertainty of numerical methods and attempt to develop some analytic understanding. The evolution of small deviations from the AdS background is governed by non-linear perturbation theory. Possibilities of significant transfer of energy between the modes (at small amplitude values) manifest themselves as secular terms (terms exhibiting unbounded growth in time) in na\"\i ve asymptotic expansions of the solutions to the equations of motion in powers of the perturbation amplitude, as noted already in \cite{Bizon:2011gg}. These terms by themselves, however, do not provide any reliable information on the ultimate fate of the system and simply signify a break-down of the na\"\i ve perturbation theory at late times. Various techniques can be employed to re-structure (resum) the na\"\i ve perturbation theory and produce modified asymptotic expansions valid at late times. Since the instability cascade takes a very long time to develop, it can only be analytically discussed in the context of such improved asymptotic expansions.

In \cite{Craps:2014vaa}, we described a perturbative resummation technique, based on the idea of the renormalization group, that produces effective equations describing the slow energy flow between the normal modes, and at the same time eliminates the secular terms at lowest non-trivial order, making the perturbation theory valid on long time intervals. A closely related technique, called the `Two-Time Framework', had been previously employed for the same system in \cite{Balasubramanian:2014cja}, though in a way geared towards numerical modelling rather than analytic study, and restricted to a finite set of low-lying modes. We have observed that only a subset of secular terms that could have appeared based on the normal mode frequency spectrum actually appear in the AdS case we study. This feature is further reflected in the effective energy flow equations we have derived, since a number of terms in those equations that could be present in fact vanish, restricting the availability of energy flow channels.

In light of the complex interplay of stability and instability that has been revealed in AdS space through numerical simulations, it is important to study precisely the constraints on the energy flow of the type we mentioned. Since the instability is generated by resonant transfer of energy to short wave-length, high-frequency modes, any limitation on the energy transfer channels available will hinder the instability onset. Such constraints are particularly apparent if formulated explicitly as conservation laws in our effective equations describing the energy transfer. This approach will form the main subject of our present study. We shall extract the three conservation laws present in the equations of \cite{Balasubramanian:2014cja,Craps:2014vaa}, analyze their origin and note that one of the three laws is explicitly related to the absence of certain types of secular terms proved in \cite{Craps:2014vaa}.

The three conservation laws we find form a direct parallel to the considerations of \cite{Basu:2014sia}, where an identical mathematical structure was described for the case of a self-interacting probe scalar field in a non-dynamical AdS space. In that paper, averaging over fast oscillations was used to produce an effective Lagrangian governing the slow energy transfer. The three conservation laws follow naturally from the symmetries of this effective Lagrangian. Since we find such perception very appealing, we shall extend this picture to the case involving fully dynamical gravity, which is algebraically much more elaborate. In addition, we shall comment on different possible implementations of the averaging over fast oscillations and the relation of this approach to the resummation schemes of \cite{Balasubramanian:2014cja,Craps:2014vaa}. This will also strengthen the theoretical foundations of the results of \cite{Basu:2014sia}.

It may appear surprising that the cases of self-interacting scalar field in a fixed background and full gravitational non-linearity appear so similar in terms of the constraints on slow energy transfer between the modes. The structure of secular terms in non-linear perturbation theory, and hence the restrictions on energy transfer channels depend crucially on the type of non-linearities involved. It happens nonetheless that scalar field self-interactions and gravitational forces produce similar energy transfer patterns in our setting.

Our observations suggest that a self-interacting scalar field in a fixed AdS background is likely to be an efficient toy model to the full gravitational weak turbulence, which is a much more complicated process. Of course, one would not be able to discuss black hole formation in this toy model setting, since the geometric background does not evolve. In general, one has to maintain a clear understanding that weak turbulence and black hole formation are distinct, even if often related, manifestations of AdS instability. Black hole formation occurs through focusing of the scalar field wave profile, for which  transfer of energy to short wavelength modes is necessary. There are settings, however, when such energy transfer occurs but a black hole does not form. A self-interacting scalar field in a fixed geometry is a completely obvious example. In a more subtle way, collapse in $\mbox{AdS}_3$ cannot occur in the weak field regime, since there is a finite minimal mass for black holes in that space. Yet, the flow of energy renders the dynamics just as turbulent as in higher dimensions \cite{Bizon:2013xha}. Turbulence is conveniently analyzed by estimating the growth of Sobolev norms (weighted sums of mode energies preferentially representing the ultraviolet modes). A classic treatment of this sort (for the weak turbulence of the non-linear Schr\"odinger equation on a torus) can be found in \cite{nls}. At the same time, as emphasized in \cite{Dimitrakopoulos:2014ada}, mode energies by themselves are insufficient for making statements about horizon formation, since focusing in position space is sensitive to phases as well as amplitudes of individual normal modes. 

The paper is organized as follows. In section \ref{pert}, we review the non-linear pertubation theory in the AdS background along the lines of \cite{Craps:2014vaa}. In section \ref{cons}, we demonstrate by a brute force verification that these equations admit three conservation laws. In section \ref{lagr}, we demonstrate that a (field-dependent) time reparametrization allows one to give the effective dynamics a Lagrangian form and relate the three conservation laws to explicit symmetries of the Lagrangian. (The reasons why this Lagrangian structure of the effective energy transfer equations only becomes apparent in certain variables are somewhat subtle and will become apparent from our subsequent systematic discussion.) We then turn to averaging methods in section \ref{ave} in hope of being able to derive a Lagrangian or Hamiltonian effective theory directly. We review the relation between averaging and the multi-scale resummation methods employed to describe the energy transfer in \cite{Balasubramanian:2014cja,Craps:2014vaa}, and we show in general how averaging can be performed directly at the level of the Hamiltonian. Finally,  in section \ref{adsave} we give a technical implementation of the averaging approach for the case of non-linear AdS perturbations.

Sections \ref{cons} and \ref{lagr} are rather technical in nature and are meant to give a matter-of-fact statement of the conservation laws in the context of the renormalization flow formalism developed in \cite{Craps:2014vaa}. A more systematic picture based on averaging methods is given in sections \ref{ave} and \ref{adsave}, which may be read semi-independently from sections \ref{cons} and \ref{lagr}.

%%%%%%%%%%%%%%%%%%%%%%%%%%%%%%%%%%%%%%%%%%%%%%%%%%%%

\section{Non-linear perturbation theory around the AdS background}
\label{pert}

\subsection{Setup of the system}

We briefly recapitulate the setup of  \cite{Bizon:2011gg, Maliborski:2013via, Craps:2014vaa} in which we will study the stability of AdS$_{d+1}$ space-time, with $d$ standing for the number of spatial dimensions. Einstein gravity with negative cosmological constant $\Lambda=-d(d-1)/(2L^{2})$ is coupled to  a free massless scalar field, leading to the equations of motion
\begin{equation}\label{Einstein}
 R_{\mu\nu}-\frac{1}{2}g_{\mu\nu}R+\Lambda g_{\mu\nu}-8\pi G\left(\partial_{\mu}\phi\partial_{\nu}\phi-\frac{1}{2}g_{\mu\nu}(\partial\phi)^{2}\right)=0
\end{equation}
and
\begin{equation}\label{scalar}
\frac{1}{\sqrt{-g}}\partial_{\mu}\left(\sqrt{-g}g^{\mu\nu}\partial_{\nu}\phi\right)=0.
\end{equation}
Restricting to spherically symmetric configurations, we consider the metric ansatz
\begin{equation}\label{eqn:MetricAnsatz}
ds^{2}=\frac{L^{2}}{\cos^{2}x}\left(\frac{dx^{2}}{A}-Ae^{-2\delta}dt^{2}+\sin^{2}x\,d\Omega_{d-1}^{2}\right),
\end{equation}
where the metric functions $A(x,t)$ and $\delta(x,t)$, as well as the scalar field $\phi(x,t)$, only depend on the time coordinate $t$, which takes values in $\mathbb{R}$, and the radial coordinate $x$, which takes values in $[0,\pi/2)$.   
The metric (\ref{eqn:MetricAnsatz}) is not completely gauge fixed: one still has the freedom to transform $\delta(x,t)\mapsto\delta(x,t)+q(t)$ together with a redefinition of the time variable $t$. Two possible gauge fixing conditions have appeared in the literature: $\delta(0,t)=0$ \cite{Bizon:2011gg, Craps:2014vaa, Balasubramanian:2014cja} and $\delta(\pi/2,t)=0$ \cite{Buchel:2014dba}. The first choice corresponds to $t$ being the proper time measured in the interior at $x=0$, while the second choice means that $t$ is the proper time measured at the boundary. 

We introduce the notation $\Phi\equiv\phi'$ and $\Pi\equiv A^{-1}e^{\delta}\dot{\phi}$ (where dots and primes denote the $t$- and $x$-derivatives, respectively) together with the convention $8\pi G=d-1$. Furthermore, it is convenient to define
\begin{equation}
\mu(x)\equiv(\tan x)^{d-1}
\qquad\text{and}\qquad
\nu(x)\equiv\frac{(d-1)}{\mu'(x)}=\frac{\sin x\cos x}{(\tan x)^{d-1}}.
\label{munu}
\end{equation}
The equations of motion then reduce to
\begin{subequations}
\label{eqn:EOM}
\begin{align}
\dot{\Phi}&=\left(Ae^{-\delta}\Pi\right)',
&\dot{\Pi}&=\frac{1}{\mu}\left(\mu Ae^{-\delta}\Phi\right)', \\
A'&=\frac{\nu'}{\nu}\left(A-1\right)-\mu\nu\left(\Phi^{2}+\Pi^{2}\right)A,
&\delta'&=-\mu\nu\left(\Phi^{2}+\Pi^{2}\right), \label{eqn:EOMConstraint}
\end{align}
\begin{equation}
\dot{A}=-2\mu\nu A^{2}e^{-\delta}\Phi\Pi.
\end{equation}
\end{subequations}
A static solution of these equations is the AdS-Schwarzschild black hole $A(x,t)=1-M\nu(x)$, $\delta(x,t)=0$ and $\phi(x,t)=0$. Unperturbed AdS space itself corresponds to $A=1$, $\delta=\phi=0$.

\subsection{Weakly non-linear perturbation theory}\label{Sec:WeaklyNonLin}

We will search for an approximate solution of the equations of motion (\ref{eqn:EOM}), subject to initial conditions $(\phi(x,t)|_{t=0},\dot{\phi}(x,t)|_{t=0})=(\epsilon\phi_{0}(x),\epsilon\psi_{0}(x))$. Therefore, we expand the unknown functions in the amplitude of the initial conditions:
\begin{equation}
\phi(x,t)=\sum_{k=0}^{\infty}\epsilon^{2k+1}\phi_{2k+1}(x,t),
\,\,\,\,\,
A(x,t)=1+\sum_{k=1}^{\infty}\epsilon^{2k}A_{2k}(x,t),
\,\,\,\,\,
\delta(x,t)=\sum_{k=1}^{\infty}\epsilon^{2k}\delta_{2k}(x,t).
\end{equation}
At first order in the $\epsilon$-expansion, the equations of motion (\ref{eqn:EOM}) are linearized and result in the homogeneous partial differential equation
\begin{equation}\label{eqn:phi1}
\ddot{\phi}_{1}+\hat{L}\phi_{1}=0
\qquad\text{with}\qquad
\hat{L}\equiv-\frac{1}{\mu(x)}\partial_{x}\left(\mu(x)\partial_{x}\right).
\end{equation}
The operator $\hat{L}$ is self-adjoint on the subspace of functions $\psi(x)$ that vanish at the boundary $\psi(\pi/2)=0$. The inner product on this Hilbert space is
\begin{equation}
\langle\psi,\chi\rangle\equiv\int_{0}^{\pi/2}\bar{\psi}(x)\chi(x)\mu(x)\text{d}x.
\end{equation}
The eigenvalues and eigenfunctions for $\hat{L}$ are $\omega_{n}^{2}$, with
\begin{equation}\label{spectr}
\omega_n=d+2n,\ \ \  n=0,1,... ,
\end{equation}
and
\begin{equation}
e_{n}(x)=k_{n}(\cos x)^{d}P_{n}^{\left(\frac{d}{2}-1,\frac{d}{2}\right)}\left(\cos(2x)\right)
\qquad\text{with}\qquad
k_{n}=\frac{2\sqrt{n!(n+d-1)!}}{\Gamma\left(n+\frac{d}{2}\right)}.
\end{equation}
The function $P_{n}^{(a,b)}(x)$ is a Jacobi polynomial of order $n$. These eigenfunctions are normalized such that $\hat{L}e_{j}=\omega_{j}^{2}e_{j}$ and $\langle e_{i},e_{j}\rangle=\delta_{ij}$. Note that all the mode frequencies $\omega_{n}$ are integer and therefore the spectrum is fully resonant. We can expand the unknown functions in the basis $\{e_{n}(x)\}$ of eigenmodes:
\begin{equation}
\phi_{2k+1}(x,t)=\sum_{n=0}^{\infty}c^{(2k+1)}_{n}(t)e_{n}(x)
\qquad\text{with}\qquad
c^{(2k+1)}_{n}(t)=\langle\phi_{2k+1}(x,t),e_{n}(x)\rangle.
\end{equation}
Equation (\ref{eqn:phi1}) then translates to $\ddot{c}_{n}^{(1)}+\omega_{n}^{2}c_{n}^{(1)}=0$ and yields the general solution of the linearized equation for $\phi_{1}$,
\begin{equation}
\phi_{1}(x,t)=\sum_{n=0}^{\infty}A_{n}\cos(\omega_{n}t+B_{n})e_{n}(x).
\end{equation}
The backreaction on the metric appears at second order. It is given by
\begin{align}
A_{2}(x,t)&=-\nu(x)\int_{0}^{x}\left(\dot{\phi}_{1}(y,t)^{2}+\phi'_{1}(y,t)^{2}\right)\mu(y)\text{d}y, \\
\delta_{2}(x,t)&=
\begin{cases}
-\int_{0}^{x}\left(\dot{\phi}_{1}(y,t)^{2}+\phi'_{1}(y,t)^{2}\right)\mu(y)\nu(y)\text{d}y & \text{ in the gauge } \delta(0,t)=0 \\
\int_{x}^{\pi/2}\left(\dot{\phi}_{1}(y,t)^{2}+\phi'_{1}(y,t)^{2}\right)\mu(y)\nu(y)\text{d}y & \text{ in the gauge } \delta(\pi/2,t)=0. \\
\end{cases}
\end{align}
At third order in the $\epsilon$-expansion, the equations of motion (\ref{eqn:EOM}) lead to the inhomogeneous equation
\begin{equation}\label{eq:S}
\ddot{\phi}_{3}+\hat{L}\phi_{3}=S\equiv2\left(A_{2}-\delta_{2}\right)\ddot{\phi}_{1}+\left(\dot{A}_{2}-\dot{\delta}_{2}\right)\dot{\phi}_{1}+\left(A'_{2}-\delta'_{2}\right)\phi'_{1}.
\end{equation}
We can project this equation onto the eigenbasis $\{e_{n}\}$, such that
\begin{equation}\label{eq:modec3}
\ddot{c}^{(3)}_{n}+\omega_{n}^{2}c^{(3)}_{n}=S_{n}
\qquad\text{with}\qquad
S_{n}=\langle S,e_{n}\rangle.
\end{equation}
After a tedious but straightforward calculation \cite{Craps:2014vaa}, one finds an explicit expression for the source term $S_{n}(t)$ in terms of the $c_{n}^{(1)}(t)$. Because the spectrum (\ref{spectr}) of linear perturbations is resonant, this source contains resonant terms that will induce secular growth of $c_{n}^{(3)}(t)$.

\subsection{Renormalization flow equations}

The secular behavior of the solutions at order $\mathcal{O}\left(\epsilon^{3}\right)$ can be resummed by absorbing it in the renormalized amplitudes $A_{l}$ and phases $B_{l}$. The renormalization group resummation of these secular terms conducted in \cite{Craps:2014vaa} (using the gauge fixing condition $\delta(0)=0$) leads to the general renormalization flow equations,
\beq\label{eqn:RGA}
\frac{2\omega_{l}}{\epsilon^{2}}\frac{dA_{l}}{d\tau}=-\underbrace{\sum_{i}^{\{i,j\}}\sum_{j}^{\neq}\sum_{k}^{\{k,l\}}}_{\omega_{i}+\omega_{j}=\omega_{k}+\omega_{l}}S_{ijkl}A_{i}A_{j}A_{k}\sin(B_{l}+B_{k}-B_{i}-B_{j})
\eeq
and
\beq\label{eqn:RGB}
\frac{2\omega_{l}A_{l}}{\epsilon^{2}}\frac{dB_{l}}{d\tau}=-T_{l}A_{l}^{3}-\sum_{i}^{i\neq l}R_{il}A_{i}^{2}A_{l}-\underbrace{\sum_{i}^{\{i,j\}}\sum_{j}^{\neq}\sum_{k}^{\{k,l\}}}_{\omega_{i}+\omega_{j}=\omega_{k}+\omega_{l}}S_{ijkl}A_{i}A_{j}A_{k}\cos(B_{l}+B_{k}-B_{i}-B_{j}).
\eeq
The coefficients that appear in these equations are written explicitly in appendix \ref{Sec:RGCoefs}. $\{i,j\}\ne\{k,l\}$ means than neither $i$ nor $j$ coincides with either $k$ or $l$. Potentially, there could have been extra contributions in these equations: terms proportional to $A_{i}A_{j}A_{k}\sin(B_{l}-B_{i}-B_{j}-B_{k})$ in (\ref{eqn:RGA}) and to $A_{i}A_{j}A_{k}\cos(B_{l}-B_{i}-B_{j}-B_{k})$ in (\ref{eqn:RGB}), from the resonant frequency addition pattern $\omega_{l}=\omega_{i}+\omega_{j}+\omega_{k}$, and terms proportional to $A_{i}A_{j}A_{k}\sin(B_{l}+B_{j}+B_{k}-B_{i})$ in (\ref{eqn:RGA}) and to $A_{i}A_{j}A_{k}\cos(B_{l}+B_{j}+B_{k}-B_{i})$ in (\ref{eqn:RGB}), from the resonant frequency addition pattern $\omega_{l}=\omega_{i}-\omega_{j}-\omega_{k}$. We have proved in  \cite{Craps:2014vaa}, however, that all such terms vanish for the AdS case. This property is not generic for all systems\footnote{For example a spherical cavity in 4-dimensional Minkowski spacetime with Dirichlet boundary conditions \cite{Maliborski:2012gx} and a holographic hard wall model in AdS$_{4}$ with Neumann boundary conditions \cite{Craps:2013iaa,Craps:2014eba} are described by equations of motion of the same form as (\ref{eqn:EOM}) and display a resonant spectrum of linearized modes. The extra terms in the renormalization flow equations vanishing in our case are present for those systems.}  that have equations of motion of the form (\ref{eqn:EOM}), but depends on the particular dynamics of AdS$_{d+1}$.

For our purposes, the symmetry properties of the non-vanishing coefficients $T_{i}$, $R_{ij}$ and $S_{ijkl}$ will be more important than their precise values. Whenever the resonance condition $\omega_{i}+\omega_{j}=\omega_{k}+\omega_{l}$ is satisfied, one has $S_{ijkl}=S_{jikl}$, $S_{ijkl}=S_{ijlk}$ and $S_{ijkl}=S_{klij}$. Another useful observation is that
\begin{equation}\label{eqn:Rij}
R_{ij}-R_{ji}=\omega_{i}^{2}(A_{jj}+\omega_{j}^{2}V_{jj})-\omega_{j}^{2}(A_{ii}+\omega_{i}^{2}V_{ii}),
\end{equation}
where the coefficients $A_{ij}$ and $V_{ij}$ are defined in appendix \ref{Sec:RGCoefs}. We can thus conclude that the $R_{ij}$ coefficients are generically non-symmetric. A noteworthy exception to this is AdS$_{3}$, as proven in appendix \ref{Sec:Rij}. 

The equations can be simplified by adopting the complex notation $\alpha_{k}=\frac{A_{k}}{2}e^{-iB_{k}}$ (used, for instance, in  \cite{Balasubramanian:2014cja}), such that the first order scalar field solution is written as
\begin{equation}\label{phi1}
\phi_{1}(x,t)=\sum_{k=0}^{\infty}A_{k}\cos(\omega_{k}t+B_{k})e_{k}(x)=\sum_{k=0}^{\infty}\left(\alpha_{k}e^{-i\omega_{k}t}+\bar{\alpha}_{k}e^{i\omega_{k}t}\right)e_{k}(x).
\end{equation}
The two renormalization flow equations (\ref{eqn:RGA}, \ref{eqn:RGB}) can then be combined into
\begin{equation}
\label{eqn:RG2}
\frac{\omega_{l}}{(2i\epsilon^{2})}\frac{d\alpha_{l}}{d\tau}=T_{l}|\alpha_{l}|^{2}\alpha_{l}+\sum_{i}^{i\,\neq\,l}R_{il}|\alpha_{i}|^{2}\alpha_{l}+\underbrace{\sum_{i}^{\{i,j\}}\sum_{j}^{\neq}\sum_{k}^{\{k,l\}}}_{\omega_{i}+\omega_{j}=\omega_{k}+\omega_{l}}S_{ijkl}\alpha_{i}\alpha_{j}\bar{\alpha}_{k}.
\end{equation}
Note that the extra terms we have described under (\ref{eqn:RGB}), which are absent due to special properties of the AdS space, would have resulted in contributions of the form $\alpha_{i}\alpha_{j}{\alpha}_{k}$, etc.\ in the above equation. This would have had an impact on the conservation laws we shall derive in section \ref{cons}.

It is instructive to define the quantity
\begin{equation}
\label{eqn:V}
V=\sum_{i}T_{i}|\alpha_{i}|^{4}+\sum_{i,j}^{i\,\neq\,j}R^{S}_{ij}|\alpha_{i}|^{2}|\alpha_{j}|^{2}+\underbrace{\sum_{i}^{\{i,j\}}\sum_{j}^{\neq}\sum_{k}^{\{k,l\}}\sum_{l}}_{\omega_{i}+\omega_{j}=\omega_{k}+\omega_{l}}S_{ijkl}\alpha_{i}\alpha_{j}\bar{\alpha}_{k}\bar{\alpha}_{l},
\end{equation}
in terms of which the renormalization flow equation (\ref{eqn:RG2}) can be simplified to
\begin{equation}
\label{eqn:RG3}
\frac{\omega_{j}}{(2i\epsilon^{2})}\frac{d\alpha_{j}}{d\tau}=\frac{1}{2}\frac{\partial V}{\partial\bar{\alpha}_{j}}+\sum_{i}R_{ij}^{A}|\alpha_{i}|^{2}\alpha_{j}.
\end{equation}
In the previous two formulas, $R_{ij}^{S}=(R_{ij}+R_{ji})/2$ and $R_{ij}^{A}=(R_{ij}-R_{ji})/2$. Note that  in (\ref{eqn:RG3}) we were allowed to drop the $i\neq j$ requirement from the sum over $i$ because $R_{ii}^{A}=0$.

%%%%%%%%%%%%%%%%%%%%%%%%%%%%%%%%%%%%%%%%%%%%%%%%%%%%

\section{Conservation laws}
\label{cons}

We now proceed to prove the existence of three conserved quantities of the renormalization flow. First note that by equation (\ref{eqn:RG2}) and its complex conjugate, we have
\begin{equation}\label{eqn:OmegaA2}
\omega_{l}\frac{d|\alpha_{l}|^{2}}{d\tau}=\omega_{l}\bar{\alpha}_{l}\frac{d\alpha_{l}}{d\tau}+\omega_{l}\alpha_{l}\frac{d\bar{\alpha}_{l}}{d\tau}=(2i\epsilon^{2})\underbrace{\sum_{i}^{\{i,j\}}\sum_{j}^{\neq}\sum_{k}^{\{k,l\}}}_{\omega_{i}+\omega_{j}=\omega_{k}+\omega_{l}}S_{ijkl}(\alpha_{i}\alpha_{j}\bar{\alpha}_{k}\bar{\alpha}_{l}-\bar{\alpha}_{i}\bar{\alpha}_{j}\alpha_{k}\alpha_{l}),
\end{equation}
and therefore
\begin{equation}
\frac{d}{d\tau}\left(\sum_{l}\omega_{l}|\alpha_{l}|^{2}\right)
=(2i\epsilon^{2})\underbrace{\sum_{i}^{\{i,j\}}\sum_{j}^{\neq}\sum_{k}^{\{k,l\}}\sum_{l}}_{\omega_{i}+\omega_{j}=\omega_{k}+\omega_{l}}S_{ijkl}(\alpha_{i}\alpha_{j}\bar{\alpha}_{k}\bar{\alpha}_{l}-\bar{\alpha}_{i}\bar{\alpha}_{j}\alpha_{k}\alpha_{l}).
\end{equation}
Since under interchange of $(i,j)\leftrightarrow(k,l)$ the coefficients $S_{ijkl}$ are symmetric, while the tensor $(\alpha_{i}\alpha_{j}\bar{\alpha}_{k}\bar{\alpha}_{l}-\bar{\alpha}_{i}\bar{\alpha}_{j}\alpha_{k}\alpha_{l})$ is antisymmetric, we find that
\begin{equation}
J=\sum_{l}\omega_{l}|\alpha_{l}|^{2}
\end{equation}
is a conserved quantity of the renormalization flow equations (\ref{eqn:RG2}). 

Similarly, from (\ref{eqn:OmegaA2}) we obtain
\begin{align}
\frac{d}{d\tau}\left(\sum_{l}\omega_{l}^{2}|\alpha_{l}|^{2}\right)
&=(2i\epsilon^{2})\underbrace{\sum_{i}^{\{i,j\}}\sum_{j}^{\neq}\sum_{k}^{\{k,l\}}\sum_{l}}_{\omega_{i}+\omega_{j}=\omega_{k}+\omega_{l}}S_{ijkl}(\alpha_{i}\alpha_{j}\bar{\alpha}_{k}\bar{\alpha}_{l}-\bar{\alpha}_{i}\bar{\alpha}_{j}\alpha_{k}\alpha_{l})\omega_{l} \nonumber \\
&=\frac{1}{2}(2i\epsilon^{2})\underbrace{\sum_{i}^{\{i,j\}}\sum_{j}^{\neq}\sum_{k}^{\{k,l\}}\sum_{l}}_{\omega_{i}+\omega_{j}=\omega_{k}+\omega_{l}}S_{ijkl}(\alpha_{i}\alpha_{j}\bar{\alpha}_{k}\bar{\alpha}_{l}-\bar{\alpha}_{i}\bar{\alpha}_{j}\alpha_{k}\alpha_{l})(\omega_{l}+\omega_{k}).
\end{align}
In the last step, we interchanged the summation indices $k$ and $l$ and used the fact that $S_{ijkl}=S_{ijlk}$. Now note that whenever the resonance condition $\omega_{i}+\omega_{j}=\omega_{k}+\omega_{l}$ is satisfied, the tensor $S_{ijkl}(\omega_{l}+\omega_{k})$ is symmetric under interchange of $(i,j)\leftrightarrow(k,l)$. Since the tensor $(\alpha_{i}\alpha_{j}\bar{\alpha}_{k}\bar{\alpha}_{l}-\bar{\alpha}_{i}\bar{\alpha}_{j}\alpha_{k}\alpha_{l})$ is antisymmetric, we find that
\begin{equation}
E=\sum_{l}\omega_{l}^{2}|\alpha_{l}|^{2}
\end{equation}
is a conserved quantity of the renormalization flow equations (\ref{eqn:RG2}). 

Finally, using the renormalization flow equations (\ref{eqn:RG3}), one can check that
\begin{align}
\frac{dV}{d\tau}&=\sum_{j}\left(\frac{\partial V}{\partial\alpha_{j}}\frac{d\alpha_{j}}{d\tau}+\frac{\partial V}{\partial\bar{\alpha}_{j}}\frac{d\bar{\alpha}_{j}}{d\tau}\right)=-\sum_{i,j}R_{ij}^{A}|\alpha_{i}|^{2}\left(\frac{2i\epsilon^{2}}{\omega_{j}}\right)\left(\bar{\alpha}_{j}\frac{\partial V}{\partial\bar{\alpha}_{j}}-\alpha_{j}\frac{\partial V}{\partial\alpha_{j}}\right) \nonumber \\
&=-2\sum_{i,j}R_{ij}^{A}|\alpha_{i}|^{2}\left(\bar{\alpha}_{j}\frac{d\alpha_{j}}{d\tau}+\alpha_{j}\frac{d\bar{\alpha}_{j}}{d\tau}\right)=-2\sum_{i,j}R_{ij}^{A}|\alpha_{i}|^{2}\frac{d|\alpha_{j}|^{2}}{d\tau}.
\end{align}
Using (\ref{eqn:Rij}) this becomes
\begin{align}
\frac{dV}{d\tau}&=-\sum_{i,j}\left(\omega_{i}^{2}(A_{jj}+\omega_{j}^{2}V_{jj})-\omega_{j}^{2}(A_{ii}+\omega_{i}^{2}V_{ii})\right)|\alpha_{i}|^{2}\frac{d|\alpha_{j}|^{2}}{d\tau} \nonumber \\
&=-E\frac{d}{d\tau}\sum_{j}(A_{jj}+\omega_{j}^{2}V_{jj})|\alpha_{j}|^{2}+\sum_{i}(A_{ii}+\omega_{i}^{2}V_{ii})|\alpha_{i}|^{2}\frac{dE}{d\tau}.
\end{align}
Since we already know that $E$ is conserved, $dE/d\tau=0$, we conclude that
\begin{equation}\label{eqn:W}
W=V+E\sum_{j}\left(A_{jj}+\omega_{j}^{2}V_{jj}\right)|\alpha_{j}|^{2}
\end{equation}
is a conserved quantity of the renormalization flow equations (\ref{eqn:RG2}). For $d=2$, this expression reduces to $W=V+E^{2}$. 

We have thus found three integrals of motion ($J$, $E$ and $W$) of the renormalization flow equations. $E$ and $W$ can be understood as the `free motion' and `interaction' energies of the oscillators comprising the scalar field. They are conserved separately under renormalization flow. $J$ is akin to a classical version of the number operator in quantum field theory (note that our normalization of $\alpha_k$ differs from the canonical normalization of creation-annihilation operators in field theory). The conservation of $J$ depends crucially on the absense of extra terms mentioned under (\ref{eqn:RGB}) and (\ref{eqn:RG2}) in the renormalization flow equations, and is specific to the AdS case we are considering.

%%%%%%%%%%%%%%%%%%%%%%%%%%%%%%%%%%%%%%%%%%%%%%%%%%%%

\section{Lagrangian form of the conservation laws}
\label{lagr}

In order to relate the conserved quantities in section \ref{cons} to symmetries using a Noether procedure, one might be tempted to try and find a Lagrangian $L(\alpha,\bar{\alpha})$ that gives rise to the renormalization flow equations (\ref{eqn:RG3}). However, since for $d\geqslant3$ the $R_{ij}$ coefficients are non-symmetric, the right hand side of (\ref{eqn:RG3}) violates an integrability condition (the curl of the force is not zero) and therefore the equations of motion cannot be derived in a usual way from a Lagrangian. In this section, we will show that this problem can be overcome by working in a different gauge for $\delta$.

\subsection{Renormalization flow equations in boundary time gauge}

The renormalization flow equations (\ref{eqn:RGA}) and (\ref{eqn:RGB}) were computed in \cite{Craps:2014vaa} using the interior time gauge fixing condition $\delta(0,t)=0$.  In section \ref{Sec:RGBoundTime}, we shall repeat that calculation in the boundary time gauge $\delta(\pi/2,t)=0$ and find the renormalization flow equations
\begin{equation}\label{eqn:RG4}
\frac{\omega_{j}}{(2i\epsilon^{2})}\frac{d\alpha_{j}}{d\tau}=\frac{1}{2}\frac{\partial W}{\partial\bar{\alpha}_{j}},
\end{equation}
where $W$ is the quantity defined in equation (\ref{eqn:W}). In contrast to the renormalization flow equations in interior time gauge (\ref{eqn:RG3}), these equations are the Euler-Lagrange equations associated to a Lagrangian,
\begin{equation}\label{eqn:EffLag}
L=\sum_{k}i\omega_{k}\left(\bar{\alpha}_{k}\frac{d\alpha_{k}}{d\tau}-\alpha_{k}\frac{d\bar{\alpha}_{k}}{d\tau}\right)+2\epsilon^{2}W.
\end{equation}
As was done in section 3.1 of \cite{Basu:2014sia} for a simpler system, we can identify three symmetries:
\begin{itemize}
\item A $U(1)$ symmetry for which all $\alpha_{n}$ have the same charge: $\alpha_{n}\mapsto e^{i\theta}\alpha_{n}$. The conserved quantity associated to this symmetry is $J=\sum_{n}\omega_{n}|\alpha_{n}|^{2}$. The absence of the possible extra terms mentioned under (\ref{eqn:RGB}) and (\ref{eqn:RG2}) is crucial for this symmetry to occur.
\item A $U(1)$ symmetry for which $\alpha_{n}$ has charge $\omega_{n}$: $\alpha_{n}\mapsto e^{i\omega_{n}\theta}\alpha_{n}$. The conserved quantity associated to this symmetry is $E=\sum_{n}\omega_{n}^{2}|\alpha_{n}|^{2}$.
\item A time translation symmetry $\tau\mapsto\tau-\tau_{0}$. The conserved quantity associated to this symmetry is $W$.
\end{itemize}
These conserved quantities are exactly the same as the ones that we determined in section \ref{cons} for the renormalization flow (\ref{eqn:RG3}) in the interior time gauge.

\subsection{Relation between the renormalization flows in different gauges}\label{relation}

One may wonder why the renormalization flow equations are derivable from a Lagrangian in one gauge but not in another. Furthermore, one may ask why the flow equations in both gauges have exactly the same conserved quantities.

In order to gain insight in these questions, we compare the result (\ref{eqn:RG4}) to the renormalization flow equations (\ref{eqn:RG3}) that appear in the interior time gauge,
\begin{equation}
\label{eqn:RG5}
\frac{\omega_{j}}{(2i\epsilon^{2})}\frac{d\beta_{j}}{d\tau}=\frac{1}{2}\frac{\partial V}{\partial\bar{\beta}_{j}}+\sum_{i}R_{ij}^{A}|\beta_{i}|^{2}\beta_{j},
\end{equation}
where we have replaced $\alpha$ by $\beta$ to highlight the difference with equation (\ref{eqn:RG4}). From the metric ansatz (\ref{eqn:MetricAnsatz}), we observe that the interior proper time $t_{I}$ and the boundary proper time $t_{B}$ are related by
\begin{align}
dt_{B}=e^{-\delta\left(\frac{\pi}{2},t_I\right)}dt_{I}&=\left(1+\epsilon^{2}\int_{0}^{\pi/2}\text{d}x\left((\varphi')^{2}+(\dot{\varphi})^{2}\right)\mu\nu+\mathcal{O}\left(\epsilon^{4}\right)\right)dt_{I}, \nonumber \\
&=\left(1+\epsilon^{2}\sum_{ij}\left(A_{ij}c_{i}c_{j}+V_{ij}\dot{c}_{i}\dot{c}_{j}\right)+\mathcal{O}\left(\epsilon^{4}\right)\right)dt_{I},
\end{align}
where, again,  the coefficients $A_{ij}$ and $V_{ij}$ are defined in appendix \ref{Sec:RGCoefs}. If one expresses $c_j$ through the complex amplitudes $\beta_j$ as $c_j=\beta_j e^{-i\om_j t_I}+\bar\beta_j e^{i\om_j t_I}$ and substitutes into the above equation, there are two types of terms: the ones rapidly oscillating (with periods of order 1) and the ones that depend on time only through slow modulations of $\beta_j$ (on time scales of order $1/\eps^2$). The former terms will only produce minuscule contributions to $t_B$ upon integration, whereas the latter can become appreciable at late times, despite being formally of order $\eps^2$. This structure is quite similar to how secular terms generally appear in perturbatively expanded solutions to the equations of motion. Retaining only the slowly varying terms, in a manner closely related to the averaging methods we shall describe in the next section, one obtains
\be
t_B\approx t_I +2\eps^2 \int^{t_I} \text{d}t \left(\sum_{i}\left(A_{ii}+\omega_{i}^{2}V_{ii}\right)|\beta_{i}|^{2}\right).
\ee 
Comparing (\ref{phi1}) in boundary and interior gauges then suggests $\alpha_j e^{-i\om_j t_B}=\beta_j e^{-i\om_j t_I}$, or
\begin{equation}\label{eqn:NonLocTrans}
\alpha_{j}(\tau)=e^{2i\epsilon^{2}\omega_{j}\theta(\tau)}\beta_{j}(\tau),
\end{equation}
with the phase
\begin{equation}
\theta(\tau)=\int_{\tau_{0}}^{\tau}\text{d}t\left(\sum_{i}\left(A_{ii}+\omega_{i}^{2}V_{ii}\right)|\beta_{i}|^{2}\right).
\end{equation}
Indeed, using the fact that this transformation satisfies $|\alpha_{i}|^{2}=|\beta_{i}|^{2}$, $W(\alpha,\bar{\alpha})=W(\beta,\bar{\beta})$ and
\begin{equation}
\frac{\omega_{j}}{(2i\epsilon^{2})}\frac{d\alpha_{j}}{d\tau}=e^{2i\epsilon^{2}\omega_{j}\theta}\left(\frac{\omega_{j}}{(2i\epsilon^{2})}\frac{d\beta_{j}}{d\tau}+\omega_{j}^{2}\beta_{j}\sum_{i}\left(A_{ii}+\omega_{i}^{2}V_{ii}\right)|\beta_{i}|^{2}\right),
\end{equation}
one can show that the transformation (\ref{eqn:NonLocTrans}) relates the renormalization flow equations (\ref{eqn:RG4}) and (\ref{eqn:RG5}) in the different gauges. This also illuminates the fact that both renormalization flow equations share the same conserved quantities.

One can check that substituting (\ref{eqn:NonLocTrans}) in the effective Lagrangian (\ref{eqn:EffLag}) leads to a local Lagrangian $L(\beta,\bar{\beta})$ despite the transformation itself being non-local. We know, however, that the renormalization flow equations in interior time gauge do not straightforwardly arise from varying a Lagrangian, so one may wonder what happens if we simply vary $L(\beta,\bar{\beta})$. First of all, if one extremizes this Lagrangian under variations that satisfy $\delta\beta|_{\tau=\tau_{i}}=\delta\beta|_{\tau=\tau_{f}}=0$ at the initial and final time, one indeed does not reproduce the renormalization flow equations. So what went wrong? In the boundary time gauge, we extremized the Lagrangian under variations that satisfied $\delta\alpha|_{\tau=\tau_{i}}=\delta\alpha|_{\tau=\tau_{f}}=0$. The point is that, because of the non-local relation (\ref{eqn:NonLocTrans}),  $\delta\alpha|_{\tau=\tau_{i}}=\delta\alpha|_{\tau=\tau_{f}}=0$ is not equivalent to $\delta\beta|_{\tau=\tau_{i}}=\delta\beta|_{\tau=\tau_{f}}=0$, but rather to a much more complicated condition that involves the values of $\delta\beta$ for all times. To summarize, in the boundary time gauge, the renormalization flow equations can be straightforwardly obtained from extremizing a Lagrangian under variations that vanish at the initial and final time. If one translates this procedure to interior time gauge, one would have to extremize the Lagrangian under variations that satisfy very unusual, awkward boundary conditions.

\subsection{Computation of the renormalization flow equations in the boundary time gauge}
\label{Sec:RGBoundTime}

We elaborate here on the computation of the renormalization flow equations in the boundary time gauge $\delta(\pi/2)=0$. Readers who are not interested in the particular details of this calculation may skip this section without loss of continuity. When repeating the calculation of \cite{Craps:2014vaa} in this new gauge, one has to replace everywhere the solution
\begin{equation}
\delta_{2}=-\int_{0}^{x}\text{d}y\left((\dot{\phi}_{1})^{2}+(\phi'_{1})^{2}\right)\mu\nu,
\end{equation}
by the expression
\begin{equation}
\delta_{2}=\int_{x}^{\pi/2}\text{d}y\left((\dot{\phi}_{1})^{2}+(\phi'_{1})^{2}\right)\mu\nu.
\end{equation}
In particular, this replacement has to be done in the computation of the source term $S_{l}$ (see appendix A of \cite{Craps:2014vaa}) that appears in the equation at order $\mathcal{O}\left(\epsilon^{3}\right)$. The only two terms that will change are $\langle\delta_{2}\ddot{\phi}_{1},e_{l}\rangle$ and $\langle\dot{\delta}_{2}\dot{\phi}_{1},e_{l}\rangle$. In the end, the source term in the boundary time gauge will be related to the source term in the interior time gauge by
\begin{align}
S_{l}^{\delta(\pi/2)=0}=S_{l}^{\delta(0)=0}&+2\sum_{i=0}^{\infty}\sum_{j=0}^{\infty}\omega_{l}^{2}c_{l}(t)\{\dot{c}_{i}(t)\dot{c}_{j}(t)V_{ij}+c_{i}(t)c_{j}(t)A_{ij}\} \nonumber \\
&-\sum_{i=0}^{\infty}\sum_{j=0}^{\infty}\dot{c}_{l}(t)\frac{\partial}{\partial t}\{\dot{c}_{i}(t)\dot{c}_{j}(t)V_{ij}+c_{i}(t)c_{j}(t)A_{ij}\}.
\end{align}
The rest of the computation is analogous to that in \cite{Craps:2014vaa} and leads to the renormalization flow equations\footnote{Effectively, we need to replace everywhere the coefficients $P_{ijl}\mapsto P_{ijl}-V_{ij}$ and $B_{ijl}\mapsto B_{ijl}-A_{ij}$.}
\begin{align}
\frac{\omega_{l}}{(2i\epsilon^{2})}\frac{d\alpha_{l}}{d\tau}&=\left(T_{l}+\omega_{l}^{2}(A_{ll}+\omega_{l}^{2}V_{ll})\right)|\alpha_{l}|^{2}\alpha_{l} \nonumber \\
&+\sum_{i}^{i\,\neq\,l}\left(R_{il}+\omega_{l}^{2}(A_{ii}+\omega_{i}^{2}V_{ii})\right)|\alpha_{i}|^{2}\alpha_{l}+\underbrace{\sum_{i}^{\{i,j\}}\sum_{j}^{\neq}\sum_{k}^{\{k,l\}}}_{\omega_{i}+\omega_{j}=\omega_{k}+\omega_{l}}S_{ijkl}\alpha_{i}\alpha_{j}\bar{\alpha}_{k}.
\end{align}
Using (\ref{eqn:W}), these equations can be written as (\ref{eqn:RG4}).

%%%%%%%%%%%%%%%%%%%%%%%%%%%%%%%%%%%%%%%%%%%%%%%%%%%%
\section{Averaging methods}
\label{ave}

In the previous section, the renormalization group equation governing the slow energy transfer between the modes due to resonant non-linear interactions were rewritten in a Lagrangian form. Thereupon, the three conservation laws restricting the energy flow became an obvious consequence of the symmetries of this effective Lagrangian. A very attractive picture for a similar set of conservation laws was obtained in \cite{Basu:2014sia} for a simpler closely related system, namely, a probe scalar field with $\phi^4$ self-interactions.

In \cite{Basu:2014sia}, an ansatz involving (linearized) fast oscillations of the scalar field with slowly drifting amplitudes and phases was substituted directly into the $\phi^4$ Lagrangian in the AdS background, and an averaging was performed over the fast oscillation, leaving an effective Lagrangian for the slow drift. This Lagrangian had a structure very similar to our (\ref{eqn:EffLag}), with the same set of symmetries and the same conservation laws,\footnote{As we explained in sections \ref{cons} and \ref{lagr}, one of the three conservation laws we find, and the corresponding Lagrangian symmetry, depend crucially on the absence of +++ and +\,-\,-  secular terms established for a free scalar field in a fully dynamical geometry in \cite{Craps:2014vaa}. It is not obvious, but true, that a similar property holds for a probe self-interacting scalar field. This can in fact be demonstrated with considerably greater ease than for the gravitational case. In short, the coefficients of the $\phi^4$ secular terms are proportional to $\int dx\, \mu\, e_i e_j e_k e_l \sec^2 x$, as can be deduced from (2.4) of  \cite{Basu:2014sia}. This can be re-written as an integral of a product of the corresponding Jacobi polynomials $P_iP_jP_kP_l$ times another fixed polynomial of degree $d-1$, with the standard Jacobi polynomial measure. If $\om_l=\om_i+\om_j+\om_k$, then $l=d+i+j+k$ and the degree of $P_l$ is higher than the sum of the degrees of all the remaining polynomials. This structure vanishes by the Jacobi polynomial orthogonality. A similar argument, with $l$ and $i$ interchanged, holds for the case $\om_l=\om_i-\om_j-\om_k$.} though of course, the exact coefficients in the potential term are different, since they depend on the exact form of the non-linearities.

It could be desirable to derive a similar elegant picture of the conservation laws for the fully dynamical gravitational instability, and we shall do that in the next section. It is worthwhile, however, to review the systematics of fast oscillation averaging and its relation to the perturbative resummation methods we have previously employed. Averaging at the level of the equation of motion is standard material in non-linear perturbation theory; a lucid and elementary detailed exposition can be found in \cite{murdock}, and will be briefly summarized here. In addition, given our interest in deriving the flow equations directly from a Lagrangian or Hamiltonian, we will develop a systematic Hamiltonian averaging method, which in the next section will be applied to our system of interest.

\subsection{Averaging over fast oscillations and the periodic normal form}

It is a natural idea that if the dynamics of a system involves rapid oscillations superimposed on slow drift behavior, there should be some sort of simplified effective description of the slow motion, in which the fast oscillations have been `integrated out'. The Born-Oppenheimer approximation in quantum mechanics is a familiar example of that.

The ideas of fast oscillation averaging in classical differential equations are well-developed and stand in close relation to methods of non-linear perturbation theory. It is not generally true that one can simply discard rapidly oscillating terms in a consistent fashion, and there are known counterexamples. However, there is a class of systems of what is known as the `periodic normal form' for which discarding rapidly oscillating terms has been proved to be accurate in a well-defined sense:
\beq
\frac{d\vec{x}}{dt}=\eps \vec{f} (\vec{x} ,t),
\label{prd}
\eeq
where $\vec{f}$ is periodic in $t$ with period $2\pi$. Here, $\vec{x}$ evolves on a timescale of order $1/\eps$, whereas $\vec{f}$ oscillates on timescales of order 1, which is fast compared to the variation of $\vec{x}$.
One can then introduce a time-averaged version of $\vec{f}$,
\beq
\vec{f}_{\mbox{\small avr}}=\frac1{2\pi}\int\limits_0^{2\pi} dt \vec{f}(\vec{x},t),
\eeq
and the corresponding averaged equation,
\beq
\frac{d\vec{x}_{\mbox{\small avr}}}{dt}=\eps \vec{f}_{\mbox{\small avr}} (\vec{x}_{\mbox{\small avr}}).
\eeq
Importantly, there is an explicit accuracy theorem for this procedure, bounding the deviations of $\vec{x}_{\mbox{\small avr}}$ from $\vec{x}$ uniformly on long time intervals. Namely, for any $T$, there exist constants $c$ and $\eps_1$ such that
\beq
|\vec{x}(t)-\vec{x}_{\mbox{\small avr}}(t)|<c\eps\qquad \mbox{for}\quad 0<t<\frac{T}\eps,\quad 0<\eps<\eps_1.
\eeq
(For a more accurate version, see theorem 6.2.2 of \cite{murdock}.)

It is straightforward to put the oscillatory systems of the sort we study,
\beq
\ddot c_j+\om_j^2c_j=S_j(c),
\label{cj}
\eeq
with a cubic\footnote{In the context of AdS, the source is a complicated non-linear function, and one works with the cubic part of its polynomial expansion at lowest order in perturbation amplitudes. We shall not discuss accurately the systematics of neglecting the higher-order terms, and simply observe that they are suppressed by higher powers of the expansion parameter.} source $S_j$ in the periodic normal form. To this end, one first introduces the conjugate momenta $\pi_j=\dot c_j$ to obtain
\beq
\dot c_j=\pi_j, \qquad \dot\pi_j=-\om_j^2c_j+S_j(c).
\label{haml}
\eeq
Note that this is just the Hamiltonian form of the equations, and we shall return to this fact in section~\ref{LagHam} in the context of the Hamiltonian averaging. One then introduces complex variables $\alpha_j(t)$ such that\footnote{For those familiar with multiscale methods, the expressions for $c_j$ and $\pi_j$ we write may look like resummed perturbative expansions including slow modulations of the complex amplitudes $\alpha_j$, truncated to the lowest order. This picture is indeed valid, and we comment on the equivalence of averaging and multiscale methods in section \ref{equiv}. Note, however, that from the onset, the averaging procedure is not formulated in the context of asymptotic expansions, but rather as a qualitatively motivated simplification in the equations of motion, which is explicitly proved to be accurate in the small $\eps$ limit.  (\ref{transfrm}) is just a change of variables, treated as exact in our context.}
\beq
c_j=\eps\left(\alpha_j e^{-i\om_j t}+\bar\alpha_j e^{i\om_j t}\right), \qquad 
\pi_j=-i\eps\om_j\left(\alpha_j e^{-i\om_j t}-\bar\alpha_j e^{i\om_j t}\right).
\label{transfrm}
\eeq
 The equations for $\alpha_j(t)$ are in the periodic normal form (we are using the fact that the source $S_j$ is a cubic polynomial in $c$):
\beq
\dot\alpha_j=\eps^2 S_j(\alpha,\bar\alpha, t),
\label{alphaexact}
\eeq
where the source has acquired an explicit periodic time dependence on $t$ through the explicit time dependences in (\ref{transfrm}). Note that the fully resonant spectrum of AdS perturbations that generates the complexity of AdS stability phenomena here works in our favor, as it makes $S_j$ in (\ref{alphaexact}) exactly periodic with a period of at most $2\pi$ in the units of AdS time we are using, since all the frequencies are integer. One can then average (\ref{alphaexact}), and the standard accuracy theorems will hold without any need for modifications. These averaged equations are in fact exactly the same as the ones describing the slow renormalization running of complex amplitudes in the context of secular term resummation, as we shall show in section~\ref{equiv}.

Note that the time-scales of order $1/\eps^2$, on which the uniform accuracy of averaging is guaranteed in (\ref{alphaexact}), are exactly the same as the time-scales for black hole formation and turbulence suggested by numerical studies. Of course, in a collapse situation, large values of fields develop in small spatial regions, invalidating our neglect of higher-order terms in the polynomial expansion of the source $S_j$ that preceeded our application of the averaging method. Nevertheless, the standard accuracy theorems give considerable strength to the averaged equations at early stages of collapse and for non-collapsing solutions. They should provide a reliable tool for probing the characteristic AdS phenomena that have been observed numerically. Note also that, in a collapse situation, there is a different form of weak field expansion that reliably describes horizon formation \cite{BM}. This latter possibility is outside our present investigation, though.

\subsection{Equivalence of averaging and multiscale methods}\label{equiv}

The averaging procedure we have described is formulated rather differently from the perturbative resummation methods we were dealing with in \cite{Craps:2014vaa} and in sections \ref{cons}-\ref{lagr} of this paper. In the context of perturbative resummation methods, one was starting with a na\"\i ve expansion of the solutions in powers of the perturbation amplitude, discovering that these expansions contained growing (secular) terms that invalidated perturbation theory at late times, and then finding a way to reorganize perturbative expansions in a way that eliminates the secular terms. This modified expansion included slow time dependences of the integration constants of the linearized (zeroth order) solutions described by renormalization flow equations very similar to the ones resulting from the averaging procedure. We would now like to see more explicitly how this happens.

One can start with the non-linear oscillator equations written in the periodic normal form (\ref{alphaexact}), construct the corresponding na\"\i ve perturbation theory, examine secular terms and see how they should be eliminated with a renormalization flow. This renormalization flow will coincide with an averaged version of (\ref{alphaexact}). (This is slightly different from the construction we employed in \cite{Craps:2014vaa}, since there, we were working in the oscillating variables of (\ref{cj}). However, those variables and the complex amplitudes $\alpha$ are related by linear transformations (\ref{transfrm}), which act straightforwardly on perturbative expansions and transform secular terms to secular terms.)

The na\"\i ve perturbative expansion for (\ref{alphaexact}) is extremely simple. The zeroth order is just $\alpha_j(t)=\alpha_{j,0}=\cnst$. One then looks for perturbative solutions of the form
\beq
\alpha_j(t)=\alpha_{j,0}+\eps^2\alpha^{(1)}_j+\cdots
\label{alpha_exp}
\eeq
$\alpha^{(1)}_j$ satisfies
\beq
\dot\alpha^{(1)}_j=S_j(\alpha_{j,0},\bar\alpha_{j,0},t),
\eeq
which is trivially solved by
\beq
\alpha^{(1)}_j(t)=\int\limits_0^t dt S_j(\alpha_{j,0},\bar\alpha_{j,0},t).
\eeq
Since $S$ is a periodic function of $t$ with a period $2\pi$, the latter expression can be written as
\beq
\alpha^{(1)}_j(t)=\frac{ t}{2\pi} \int\limits_0^{2\pi} dt S_j(\alpha_{j,0},\bar\alpha_{j,0},t)+\alpha^{(1,\mbox{\small non-secular})}_j(t),
\label{alpha1}
\eeq
where $\alpha^{(1,\mbox{\small non-secular})}_j(t)$ remains bounded at large times and does not compromise the validity of perturbation theory.

The renormalization method (like other related multi-scale methods) gives a prescription for eliminating the first (secular) term on the right-hand side of (\ref{alpha1}) that grows with time and invalidates na\"\i ve perturbation theory at times of order $1/\eps^2$. (We only give a brief practical sketch here; further details and explanations can be found in \cite{Craps:2014vaa}.) Given a secular term proportional to $t$, one introduces an arbitrary time $\tau$, writes $t=(t-\tau)+\tau$ and absorbs the contribution proportional to $\tau$ in `renormalized' integration constants of the zeroth order solution. In our case, the integration constants $\alpha_{j,0}$ are related to their renormalized versions by
\beq
\alpha_{j,0}=\alpha_{j,R}(\tau)- \frac{\eps^2 \tau}{2\pi} \int\limits_0^{2\pi} dt S_j(\alpha_{j,R}(\tau),\bar\alpha_{j,R}(\tau),t).
\eeq
If this is substituted into (\ref{alpha_exp}), the secular term in (\ref{alpha1}) is indeed exactly cancelled at the moment $t=\tau$. One then demands that the expansions with different `renormalization scale' $\tau$ represent the same solution, 
\beq
\frac{d}{d\tau}\left[\alpha_{j,R}(\tau)+ \frac{\eps^2 (t-\tau)}{2\pi} \int\limits_0^{2\pi} dt S_j(\alpha_{j,R}(\tau),\bar\alpha_{j,R}(\tau),t)\right]=0.
\eeq
To leading nontrivial order in $\epsilon$, this results in
\beq
\frac{d\alpha_{j,R}}{d\tau}=\frac{\eps^2}{2\pi} \int\limits_0^{2\pi} dt S_j(\alpha_{j,R}(\tau),\bar\alpha_{j,R}(\tau),t),
\eeq
exactly identical to the averaged form of (\ref{alphaexact}). Finally, one sets $\tau=t$. By the standard lore of renormalization, perturbation theory expressed through $\alpha_{j,R}(\tau)$ at a gliding scale $\tau=t$ is free of (lowest order) secular terms at all times and valid on long time intervals. Since we have established equivalence of the lowest order renormalization resummation and lowest order averaging, the standard accuracy theorems for averaging also apply.

\subsection{Lagrangian and Hamiltonian averaging}\label{LagHam}

One of the primary motivations for our appeal to averaging methods has been the picture of conservation laws for a probe self-interacting scalar field developed in \cite{Basu:2014sia}. By performing averaging directly in the Lagrangian, the authors derive a Lagrangian form of the effective theory as a descendant of the Lagrangian form of the fundamantal theory. The conservation laws are manifest in this procedure. By contrast, in multi-scale resummation approaches, the effective equations for slow energy transfer are derived by specific techniques having nothing to do with the Lagrangian formalism, and the Lagrangian nature of the resulting flow equations has to be guessed, together with the conservation laws.

Even though the qualitative picture developed in \cite{Basu:2014sia} is very attractive, the practical implementation of averaging can be considerably improved in terms of consistency and rigor. The authors consider a Lagrangian for oscillators with weak non-linear couplings (the same structure that we are dealing with, only the values of the couplings are different for their system) and make the following substitution (more similar in spirit to the `Two-Time Framework' than to rigorous implementations of averaging):
\beq
c_j=\alpha_j(\tau)e^{i\om_j t}+\bar\alpha_j(\tau)e^{-i\om_j t}.
\eeq
They then average the Lagrangian over time to be left with a theory for $\tau$-evolution. It is difficult to implement this procedure accurately however, since one sometimes pretends that $\tau$-dependent (but $t$-independent) terms are constants, but while differentiating $c_j$, one uses $\tau=\eps^2 t$, as in the `Two-Time Framework' of \cite{Balasubramanian:2014cja} (remember that the expansion parameter is $\eps^2$ for the AdS case). At the same time, terms containing two $\tau$-derivatives are discarded on the basis of being `small', whereas discarding highest derivative terms is, in general, a subtle operation. The results of \cite{Basu:2014sia} are essentially correct and must be derivable by more accurate methods. We shall now explain how to bring them in accord with the standard lore of fast oscillation averaging.

Since most rigorous results on averaging are formulated in the context of the first order periodic normal form equations (\ref{prd}), it is natural to work in the Hamiltonian, rather than Lagrangian formalism, since the Hamiltonian equations are naturally first order. Of course, if one obtains an effective Hamiltonian theory for the slow energy transfer at the end, it is straightforward to convert it to a Lagrangian theory. Furthermore, conservation laws can be deduced from symmetries of the Hamiltonian directly.

One can easily bring the equations for a system with a Hamiltonian $H_0+\eps H_1$ to the periodic normal form (\ref{prd}) while maintaining their Hamiltonian character. This can be accomplished in a number of ways, for example by the following method, which is the classical analog of the familiar quantum-mechanical `interaction picture'. Let $q$ and $p$ be the canonical coordinates and momenta of the original system. One can then define new (`interaction picture') canonical variables $\tilde q$ and $\tilde p$ by the following (time-dependent) canonical transformation: for given $q$ and $p$ at moment $t$, we define $\tilde q$ and $\tilde p$ to be the initial conditions at moment 0 that, under the evolution induced by $H_0$, evolve to $q$ and $p$ at moment $t$. Such transformations induced by a Hamiltonian evolution are known to be canonical, with a generating function equal to the action $S_0$ of the classical solution of $H_0$ connecting $\tilde q$ at $t=0$ with $q$ at $t$. The new Hamiltonian for $\tilde q$, $\tilde p$ is
\beq
\tilde H=H_0+\eps H_1+\frac{\del S_0}{\del t}=\eps H_1,
\label{Htilde}
\eeq
where we have used the Hamilton-Jacobi equation for $H_0$, and the equations of motion are explicitly of the periodic normal form:
\beq
\frac{d\tilde p}{dt}=-\eps\frac{\del H_1}{\del\tilde q},\qquad \frac{d\tilde q}{dt}=\eps\frac{\del H_1}{\del \tilde p}.
\label{Hamlint}
\eeq
Since the canonical transformation we have employed depends on time, $H_1$ expressed in terms of the new variables also has an explicit dependence on time. Averaging over that dependence commutes with differentiation with respect to $\tilde q$ and $\tilde p$, hence the result of applying the standard averaging procedure would still be in the Hamiltonian form,
\beq
\frac{d\tilde p}{dt}=-\eps\frac{\del\bar H}{\del\tilde q},\qquad \frac{d\tilde q}{dt}=\eps\frac{\del\bar H}{\del \tilde p},
\eeq
where $\bar H$ is the time average of $H_1$ (after it has been expressed in the `interaction picture', and only averaging over explicit time dependences is understood, as in all the rigorous implementations of averaging we have described). Thus, the standard averaging of the equations of motion in the `periodic normal form' can be simply implemented by averaging (explicit time dependence of) the Hamiltonian expressed in the `interaction picture'.

In application to systems of non-linear oscillators of the form (\ref{cj}), we shall simply employ the transformation (\ref{transfrm}), closely related to the canonical transformation described above (\ref{Htilde}). The only difference is that in addition to cancelling the part of the evolution corresponding to the free part of the Hamiltonian, one changes to the complex amplitude representation. The resulting equations of motion are in the periodic normal form, being the Hamiltonian version of (\ref{alphaexact}). Similarly to what we described above, one can apply the averaging procedure to these equations by simply applying it to the Hamiltonian. This is a straightforward reformulation of the standard averaging and the validity of the accuracy theorems is maintained.

%%%%%%%%%%%%%%%%%%%%%%%%%%%%%%%%%%%%%%%%%%%%%%%%%%%%

\section{AdS averaging}
\label{adsave}

\subsection{Effective action for the scalar field}

Since our aim is to obtain an effective averaged Hamiltonian theory for the slow energy transfer between the scalar field modes, we must start by revealing the Lagrangian/Hamiltonian structures in the underlying fundamental theory.
The field equations (\ref{Einstein}) and (\ref{scalar}) can be reproduced by extremizing the action
 \begin{equation}\label{eqn:Action}
\mathcal{S}=\int_{\mathcal{M}}\text{d}^{d+1}\mathbf{x}\sqrt{-g}\left(\frac{1}{16\pi G}(R-2\Lambda)-\frac{1}{2}(\partial\phi)^{2}\right)+\frac{1}{8\pi G}\int_{\partial\mathcal{M}}\text{d}^{d}\mathbf{x}\sqrt{-\gamma}K+\mathcal{S}_{C},
\end{equation}
where the boundary term consists of the Gibbons-Hawking term and a holographic counterterm
\begin{equation}
\mathcal{S}_{C}=-\frac{1}{8\pi G}\int_{\partial\mathcal{M}}\text{d}^{d}\mathbf{x}\sqrt{-\gamma}\left(\frac{d-1}{L}\right).
\end{equation}
We can write the variation of this action as
\begin{align}\label{variation}
\delta\mathcal{S}
&=\int_{\mathcal{M}}\text{d}^{d+1}\mathbf{x}\sqrt{-g}\left(\frac{1}{16\pi G}E_{\mu\nu}\delta g^{\mu\nu}+\Box\phi\delta\phi\right) \nonumber \\
&+\int_{\partial\mathcal{M}}\text{d}^{d}\mathbf{x}\sqrt{-\gamma}\left(\frac{1}{16\pi G}\left(K_{ij}-K\gamma_{ij}+\left(\frac{d-1}{L}\right)\gamma_{ij}\right)\delta\gamma^{ij}-n^{\mu}\partial_{\mu}\phi\delta\phi\right),
\end{align}
where $E_{\mu\nu}$ and $\Box\phi$ are the left-hand sides of (\ref{Einstein}) and (\ref{scalar}).  Demanding that this should vanish under variations which at the boundary satisfy $\delta\gamma^{ij}=0$ and $\delta\phi=0$, indeed leads to (\ref{Einstein}) and (\ref{scalar}).

Using the constraint equations (\ref{eqn:EOMConstraint}), one can integrate out the metric dependence in the action (\ref{eqn:Action}) and expand to lowest non-trivial order in powers of the scalar field. The constraint equations (\ref{eqn:EOMConstraint}) can be rewritten as
\begin{equation}
1-A=e^{\delta}\nu\int_{0}^{x}\text{d}y\left(\Phi^{2}+\Pi^{2}\right)e^{-\delta}\mu
\end{equation}
and
\begin{equation}
\delta=\int_{x}^{\pi/2}\text{d}y\left(\Phi^{2}+\Pi^{2}\right)\mu\nu,
\end{equation}
and can be solved perturbatively in powers of the scalar field $\phi$. Substituting the resulting expressions for the metric functions in the action, one obtains an effective action for the scalar field. For the boundary time gauge $\delta(\pi/2,t)=0$, the computation described in appendix \ref{Sec:EffAction} results in the effective action
\begin{equation}
\tilde{S}=-\frac{L^{d-1}V_{d-1}}{2}\int\text{d}x\int\text{d}t\left(\left((\phi')^{2}-(\dot{\phi})^{2}\right)\mu-\left((\phi')^{2}+(\dot{\phi})^{2}\right)\mu\nu\int_{0}^{x}\text{d}y\left((\phi')^{2}+(\dot{\phi})^{2}\right)\mu\right),
\end{equation}
where $V_{d-1}$ represents the volume of the sphere $\mathcal{S}^{d-1}$. This gives an effective action for $\phi$ up to first non-trivial order in the interactions.

We briefly pause to discuss why we computed the effective action in boundary time gauge rather than interior time gauge. One can anticipate as follows that this is the correct choice. After solving for the metric functions using the constraints, it is clear that the variations in (\ref{variation}) are no longer independent, since, in any given gauge, the metric functions are specific functionals of the scalar field. If we consider variations of the scalar field that vanish at the boundary, then in boundary time gauge the metric variations vanish automatically at the boundary (e.g., the metric function $\delta$ vanishes at the boundary by the gauge condition, and so will its variation) and the variational principle straightforwardly reproduces the correct equations of motion. In interior time gauge, however,  variations of the scalar field that vanish at the boundary will generically lead to variations of the metric function $\delta$ that do not vanish at the boundary (because the boundary value of $\delta$ also depends on the scalar field in the bulk), so a naive implementation of the variational principle would not reproduce the correct equations of motion. This difference between the two gauges is similar and related to the difference discussed at the end of section~\ref{relation} for renormalization flows, where a useful Lagrangian description could only be found in the boundary time gauge. We will therefore limit our attention to the boundary time gauge.

Expanding in modes $\phi(x,t)=\epsilon\sum_{k}c_{k}(t)e_{k}(x)$ and defining the coefficients
\begin{equation}\label{eqn:WCoeffs}
W_{ijkl}^{(a,b)}=\int_{0}^{\pi/2}\text{d}x\,e^{(a)}_{i}(x)e^{(a)}_{j}(x)\mu(x)\nu(x)\int_{0}^{x}e^{(b)}_{k}(y)e^{(b)}_{l}(y)\mu(y),
\end{equation}
where $e^{(a)}_{i}$ denotes the $a$th derivative of $e_{i}$, etc,
we can write this effective action as $\tilde{S}=L^{d-1}V_{d-1}\epsilon^{2}\int\text{d}t\,\mathcal{L}$, with  Lagrangian
\begin{align}\label{eqn:EffLag2}
\mathcal{L}&=\frac{1}{2}\sum_{k}\left(\dot{c}_{k}^{2}-\omega_{k}^{2}c_{k}^{2}\right) \\
&+\frac{\epsilon^{2}}{2}\sum_{ijkl}\left(c_{i}c_{j}c_{k}c_{l}W^{(1,1)}_{(ijkl)}+\dot{c}_{i}\dot{c}_{j}c_{k}c_{l}W^{(0,1)}_{ijkl}+c_{i}c_{j}\dot{c}_{k}\dot{c}_{l}W^{(1,0)}_{ijkl}+\dot{c}_{i}\dot{c}_{j}\dot{c}_{k}\dot{c}_{l}W^{(0,0)}_{(ijkl)}\right)+\mathcal{O}\left(\epsilon^{4}\right). \nonumber
\end{align}
Because of the interchange symmetry of the arguments, we really need the symmetric part
\begin{equation}
W_{(ijkl)}^{(a,a)}=\frac{1}{6}\left(W_{ijkl}^{(a,a)}+W_{klij}^{(a,a)}+W_{ikjl}^{(a,a)}+W_{jlik}^{(a,a)}+W_{iljk}^{(a,a)}+W_{jkil}^{(a,a)}\right).
\end{equation}

\subsection{Averaged Hamiltonian system}
\label{Sec:AveragedHamiltonian}

For the effective Lagrangian (\ref{eqn:EffLag2}), the canonical momenta are given by
\begin{align}
\pi_{k}=\frac{\partial\mathcal{L}}{\partial\dot{c}_{k}}=&\dot{c}_{k}+\epsilon^{2}\sum_{ijl}\left(\dot{c}_{l}c_{i}c_{j}W^{(0,1)}_{klij}+c_{i}c_{j}\dot{c}_{l}W^{(1,0)}_{ijkl}+2\dot{c}_{i}\dot{c}_{j}\dot{c}_{l}W^{(0,0)}_{(ijkl)}\right)+\mathcal{O}\left(\epsilon^{4}\right),
\end{align}
such that the Hamitonian $\mathcal{H}=\sum_{k}\pi_{k}\dot{c}_{k}-\mathcal{L}$ becomes
\begin{align}
\label{mathcalH}
\mathcal{H}&=\frac{1}{2}\sum_{k}\left(\pi_{k}^{2}+\omega_{k}^{2}c_{k}^{2}\right) \\
&-\frac{\epsilon^{2}}{2}\sum_{ijkl}\left(c_{i}c_{j}c_{k}c_{l}W^{(1,1)}_{(ijkl)}+\pi_{i}\pi_{j}c_{k}c_{l}W^{(0,1)}_{ijkl}+c_{i}c_{j}\pi_{k}\pi_{l}W^{(1,0)}_{ijkl}+\pi_{i}\pi_{j}\pi_{k}\pi_{l}W^{(0,0)}_{(ijkl)}\right)+\mathcal{O}\left(\epsilon^{4}\right). \nonumber
\end{align}
Next, one performs a (time dependent) canonical transformation of the sort we described above (\ref{Htilde}):
\beq
c_k=\tilde c_k\cos\om_k t+\frac{\tilde\pi_k}{\om_k}\sin\om_k t,\qquad\pi_k=\tilde\pi_k\cos\om_k t-\om_k\tilde c_k\sin\om_k t.
\eeq
The time dependences are chosen precisely in a way that puts the system in the `interaction picture', i.e., cancels the free evolution given by the first line of (\ref{mathcalH}). The new equations of motion are
\begin{equation}
\dot{\tilde{c}}_{k}=\frac{\partial\mathcal{\tilde H}}{\partial\tilde\pi_{k}},\qquad \dot{\tilde\pi}_{k}=-\frac{\partial\mathcal{\tilde H}}{\partial\tilde c_{k}}.
\label{alphaHaml}
\end{equation}
where $\mathcal{\tilde H}$ is the second line of (\ref{mathcalH}), expressed through $\tilde c_k$ and $\tilde\pi_k$. 

Equations (\ref{alphaHaml}) are of the periodic normal form. As per our general discussion in section \ref{ave}, the standard averaging procedure can be implemented by simply averaging the explicit time dependence in $\mathcal{\tilde H}$ (acquired due to the explicit time dependence of the canonical transformation we have employed):
\begin{equation}
\overline{\mathcal{H}}(\tilde c_k,\tilde\pi_k)=\frac{1}{2\pi}\int\limits_{0}^{2\pi}\text{d}t\,\mathcal{\tilde H}(\tilde c_k,\tilde\pi_k,t).
\label{Hamlave}
\end{equation}
It is convenient to re-express the averaged Hamiltonian through the complex amplitudes $\alpha_k$ that we have been using in the preceeding sections of this paper:
\beq
\tilde c_{k}=\alpha_{k}+\bar{\alpha}_{k}, \qquad
\tilde\pi_{k}=-i\omega_{k}(\alpha_{k}-\bar{\alpha}_{k}).
\eeq
Since this transformation is time-independent, it does not interfere with the averaging in (\ref{Hamlave}):
\begin{equation}
\overline{\mathcal{H}}(\alpha_{k},\bar{\alpha}_{k})=\frac{1}{2\pi}\int\limits_{0}^{2\pi}\text{d}t\,\mathcal{\tilde H}(\alpha_{k},\bar{\alpha}_{k},t).
\label{avealpha}
\end{equation}
The relation of the original $c_k, \pi_k$ and the complex amplitudes $\alpha_k$ is given by the standard formulas,
\beq
c_{k}=e^{-i\omega_{k}t}\alpha_{k}+e^{i\omega_{k}t}\bar{\alpha}_{k}, \qquad
\pi_{k}=-i\omega_{k}\left(e^{-i\omega_{k}t}\alpha_{k}-e^{i\omega_{k}t}\bar{\alpha}_{k}\right).
\eeq
The averaged form of (\ref{alphaHaml}) is simply
\beq
\dot{\tilde{c}}_{k}=\frac{\partial\overline{\mathcal{H}}}{\partial\tilde\pi_{k}},\qquad \dot{\tilde\pi}_{k}=-\frac{\partial\overline{\mathcal{H}}}{\partial\tilde c_{k}},
\eeq
or in terms of the complex amplitudes
\beq
\dot{\alpha}_{k}=\frac{1}{2i\omega_{k}}\frac{\partial\overline{\mathcal{H}}}{\partial\bar{\alpha}_{k}}.
\label{alphadot}
\eeq
 After some algebra, described in detail in appendix \ref{Sec:CoefsRGvsAve}, one finds that computing (\ref{avealpha}) gives
\begin{equation}\label{eqn:AveHam}
\overline{\mathcal{H}}=-2\epsilon^{2}W,
\end{equation}
where $W$ is the quantity defined in (\ref{eqn:W}). Equation (\ref{alphadot}) then becomes
\begin{equation}
-i\omega_{k}\dot{\alpha}_{k}=\epsilon^{2}\frac{\partial W}{\partial\bar{\alpha}_{k}},
\end{equation}
which is exactly the renormalization flow equations (\ref{eqn:RG4}). The averaged Lagrangian $\overline{\mathcal{L}}$ corresponding to the averaged Hamiltonian $\overline{\mathcal{H}}$ is exactly the Lagrangian (\ref{eqn:EffLag}) that appeared in section \ref{lagr}. The conservation laws, which have been the main subject-matter of our treatment, are straightforward consequences of the symmetries of these averaged Lagrangian and Hamiltonian, as described under  (\ref{eqn:EffLag}).

%%%%%%%%%%%%%%%%%%%%%%%%%%%%%%%%%%%%%%%%%%%%%%%%%%%%

\section{Acknowledgments}

We would like to thank Pallab Basu, Alex Buchel, Chethan Krishnan, Gautam Mandal and Shiraz Minwalla for useful discussions.
The work of B.C.\ and J.V.\ has been supported by the Belgian Federal Science Policy Office through the Interuniversity Attraction Pole P7/37, by FWO-Vlaanderen through project G020714N, and by the Vrije Universiteit Brussel through the Strategic Research Program ``High-Energy Physics.'' The research of O.E.\ has been supported by Ratchadaphisek Sompote Endowment Fund. J.V.\ is supported by a PhD Fellowship of the Research Foundation Flanders (FWO).

%%%%%%%%%%%%%%%%%%%%%%%%%%%%%%%%%%%%%%%%%%%%%%%%%%%%

\appendix

\section{Coefficients of the renormalization flow equations}
\label{Sec:RGCoefs}

Here, we shall summarize (simplified versions of) the expressions of \cite{Craps:2014vaa} for the coefficients that appear in the renormalization flow equations in terms of integrals of the mode functions,\footnote{Compared to the corresponding results in \cite{Craps:2014vaa}, we used that $H_{ijkl}=\omega_{k}^{2}X_{ijkl}-Y_{ijkl}+\omega_{i}^{2}X_{klij}-Y_{klij}$ and $M_{ijk}=\omega_{i}^{2}W_{ijk}+B_{ijk}-A_{ij}-X_{ijkk}$, which can be shown using integration by parts, along with the relations $W_{ijk}=W_{ijkk}$, $P_{ijk}=V_{ij}-W_{ijkk}$ and $B_{ijk}=A_{ij}-W^{*}_{ijkk}$, which follow directly from the definitions.}
\begin{align}
T_{l}=&\frac{1}{2}\omega_{l}^{2}X_{llll}+\frac{3}{2}Y_{llll}+2\omega_{l}^{4}W_{llll}+2\omega_{l}^{2}W^{*}_{llll}-\omega_{l}^{2}(A_{ll}+\omega_{l}^{2}V_{ll}), \\
R_{il}=&
\frac{1}{2}\left(\frac{\omega_{i}^{2}+\omega_{l}^{2}}{\omega_{l}^{2}-\omega_{i}^{2}}\right)\left(\omega_{l}^{2}X_{illi}-\omega_{i}^{2}X_{liil}\right)+2\left(\frac{\omega_{l}^{2}Y_{ilil}-\omega_{i}^{2}Y_{lili}}{\omega_{l}^{2}-\omega_{i}^{2}}\right)+\left(\frac{\omega_{i}^{2}\omega_{l}^{2}}{\omega_{l}^{2}-\omega_{i}^{2}}\right)\left(X_{illi}-X_{lili}\right) \nonumber \\
&+\frac{1}{2}(Y_{iill}+Y_{llii})+\omega_{i}^{2}\omega_{l}^{2}(W_{llii}+W_{iill})+\omega_{i}^{2}W^{*}_{llii}+\omega_{l}^{2}W^{*}_{iill}-\omega_{l}^{2}(A_{ii}+\omega_{i}^{2}V_{ii}),\\
%\end{align}
%\begin{align}
S_{ijkl}=&
-\frac{1}{4}\left(\frac{1}{\omega_{i}+\omega_{j}}+\frac{1}{\omega_{i}-\omega_{k}}+\frac{1}{\omega_{j}-\omega_{k}}\right)(\omega_{i}\omega_{j}\omega_{k}X_{lijk}-\omega_{l}Y_{iljk}) \nonumber \\
&-\frac{1}{4}\left(\frac{1}{\omega_{i}+\omega_{j}}+\frac{1}{\omega_{i}-\omega_{k}}-\frac{1}{\omega_{j}-\omega_{k}}\right)(\omega_{j}\omega_{k}\omega_{l}X_{ijkl}-\omega_{i}Y_{jikl}) \nonumber \\
&-\frac{1}{4}\left(\frac{1}{\omega_{i}+\omega_{j}}-\frac{1}{\omega_{i}-\omega_{k}}+\frac{1}{\omega_{j}-\omega_{k}}\right)(\omega_{i}\omega_{k}\omega_{l}X_{jikl}-\omega_{j}Y_{ijkl}) \nonumber \\
&-\frac{1}{4}\left(\frac{1}{\omega_{i}+\omega_{j}}-\frac{1}{\omega_{i}-\omega_{k}}-\frac{1}{\omega_{j}-\omega_{k}}\right)(\omega_{i}\omega_{j}\omega_{l}X_{kijl}-\omega_{k}Y_{ikjl}),\\
%\end{align}
%and
%\begin{align}
Q_{ijkl}=&\hspace{3mm}
\frac{1}{12}\left(\frac{1}{\omega_{i}+\omega_{j}}+\frac{1}{\omega_{i}+\omega_{k}}+\frac{1}{\omega_{j}+\omega_{k}}\right)(\omega_{i}\omega_{j}\omega_{k}X_{lijk}+\omega_{l}Y_{iljk}) \nonumber \\
&+\frac{1}{12}\left(\frac{1}{\omega_{i}+\omega_{j}}+\frac{1}{\omega_{i}+\omega_{k}}-\frac{1}{\omega_{j}+\omega_{k}}\right)(\omega_{j}\omega_{k}\omega_{l}X_{ijkl}+\omega_{i}Y_{jikl}) \nonumber \\
&+\frac{1}{12}\left(\frac{1}{\omega_{i}+\omega_{j}}-\frac{1}{\omega_{i}+\omega_{k}}+\frac{1}{\omega_{j}+\omega_{k}}\right)(\omega_{i}\omega_{k}\omega_{l}X_{jikl}+\omega_{j}Y_{ijkl}) \nonumber \\
&+\frac{1}{12}\left(-\frac{1}{\omega_{i}+\omega_{j}}+\frac{1}{\omega_{i}+\omega_{k}}+\frac{1}{\omega_{j}+\omega_{k}}\right)(\omega_{i}\omega_{j}\omega_{l}X_{kijl}+\omega_{k}Y_{ikjl}),
\end{align}
where $Q_{ijkl}$ is the would-be coefficient of the +++ secular terms, proved to vanish in \cite{Craps:2014vaa}. The expressions for the $S$ and $Q$ coefficients given above are substantial simplifications of what has been published previously.
The integrals that appear in these expressions are defined by
\begin{subequations}
\begin{align}
X_{ijkl}&=\int_{0}^{\pi/2}\text{d}x\,e'_{i}(x)e_{j}(x)e_{k}(x)e_{l}(x)(\mu(x))^{2}\nu(x), \\
Y_{ijkl}&=\int_{0}^{\pi/2}\text{d}x\,e'_{i}(x)e_{j}(x)e'_{k}(x)e'_{l}(x)(\mu(x))^{2}\nu(x), 
\end{align}\vspace{5mm}
\begin{align}
W_{ijkl}&=\int_{0}^{\pi/2}\text{d}x\,e_{i}(x)e_{j}(x)\mu(x)\nu(x)\int_{0}^{x}\text{d}y\,e_{k}(y)e_{l}(y)\mu(y), \\
W^{*}_{ijkl}&=\int_{0}^{\pi/2}\text{d}x\,e'_{i}(x)e'_{j}(x)\mu(x)\nu(x)\int_{0}^{x}\text{d}y\,e_{k}(y)e_{l}(y)\mu(y), \\
V_{ij}&=\int_{0}^{\pi/2}\text{d}x\,e_{i}(x)e_{j}(x)\mu(x)\nu(x), \\
A_{ij}&=\int_{0}^{\pi/2}\text{d}x\,e'_{i}(x)e'_{j}(x)\mu(x)\nu(x).
\end{align}
\end{subequations}

\section{Symmetry of $R_{ij}$ coefficients in AdS$_{3}$}
\label{Sec:Rij}

Renormalization coefficients possess some special properties for $d=2$. We shall now show that $R_{ij}=R_{ji}$ in AdS$_{3}$. First, we introduce
\begin{equation}
V_{ij}=\int_{0}^{\frac{\pi}{2}}\text{d}x\,e_{i}e_{j}\mu\nu
\quad\text{,}\quad
N_{ij}=\int_{0}^{\frac{\pi}{2}}\text{d}x\,e'_{i}e_{j}\mu\nu'.
\quad\text{and}\quad
A_{ij}=\int_{0}^{\frac{\pi}{2}}\text{d}x\,e'_{i}e'_{j}\mu\nu.
\end{equation}
Using integration by parts and the fact that $(\mu e'_{i})'=-\omega_{i}^{2}\mu e_{i}$ and $(\mu\nu')'=-4\mu\nu$, we find that
\begin{equation}
N_{ij}+A_{ij}=\omega_{i}^{2}V_{ij}
\end{equation}
and
\begin{equation}
N_{ij}+N_{ji}=C_{i}C_{j}+4V_{ij}
\qquad\text{with}\qquad
C_{i}\equiv\frac{2\sqrt{d-2}}{\Gamma(d/2)}\sqrt{\frac{(i+d-1)!}{i!}}.
\end{equation}
For this last result, we have also used that $\lim_{x\rightarrow0}\left(e_{i}(x)e_{j}(x)\mu(x)\nu'(x)\right)=-C_{i}C_{j}$. Combining these two equations, we find that
\begin{equation}
R_{ij}-R_{ji}=\omega_{i}^{2}(A_{jj}+\omega_{j}^{2}V_{jj})-\omega_{j}^{2}(A_{ii}+\omega_{i}^{2}V_{ii})=2\omega_{i}^{2}(\omega_{j}^{2}-1)V_{jj}-2\omega_{j}^{2}(\omega_{i}^{2}-1)V_{ii}-\frac{\omega_{i}^{2}}{2}C_{j}^{2}+\frac{\omega_{j}^{2}}{2}C_{i}^{2}.
\end{equation}
We can use the identity
\begin{equation}
(2n+a+b+2)(1+x)P_{n}^{(a,b+1)}(x)=(2n+2b+2)P_{n}^{(a,b)}(x)+(2n+2)P_{n+1}^{(a,b)}(x),
\end{equation}
to rewrite the eigenfunctions for $d=2$ as
\begin{equation}
e_{n}(x)=k_{n}(\cos x)^{2}P_{n}^{\left(0,1\right)}\left(\cos(2x)\right)=\frac{1}{2}k_{n}\left(P_{n}^{\left(0,0\right)}\left(\cos(2x)\right)+P_{n+1}^{\left(0,0\right)}\left(\cos(2x)\right)\right).
\end{equation}
We can relate this to associated Legendre polynomials $p^{m}_{l}(x)$ as
\begin{equation}
P_{n}^{(a,a)}(x)=(-1)^{a}\frac{2^{a}(a+n)!}{(2a+n)!}\left(1-x^{2}\right)^{-\frac{a}{2}}p^{a}_{a+n}(x),
\end{equation}
such that
\begin{equation}
e_{n}(x)=\sqrt{n+1}\left(p_{n}(\cos(2x))+p_{n+1}(\cos(2x))\right),
\end{equation}
In this expression, the functions $p_{n}(x)=p^{0}_{n}(x)$ are the ordinary Legendre polynomials. These are defined by
\begin{equation}
p_{n}(x)=\frac{1}{2^{n}n!}\frac{d^{n}}{dx^{n}}\left[\left(x^{2}-1\right)^{n}\right],
\end{equation}
and satisfy the following useful identity,
\begin{equation}
\int_{0}^{\pi/2}p_{n}(\cos(2x))p_{m}(\cos(2x))\sin x\cos x\,\text{d}x=\frac{1}{2(2n+1)}\delta_{nm}.
\end{equation}
It follows from this expression that for $d=2$, we have that $V_{ii}=\frac{1}{2}\left(\frac{\omega_{i}^{2}}{\omega_{i}^{2}-1}\right)$. Notice that also $C_{i}=0$ for $d=2$. This completes the proof that $R_{ij}=R_{ji}$ for $d=2$.

\section{Effective action for the scalar field}
\label{Sec:EffAction}

An effective action for the scalar field can be obtained by integrating out the metric components from the action (\ref{eqn:Action}) using the constraint equations (\ref{eqn:EOMConstraint}). If we formally extract the amplitude of the scalar field as $\phi=\xi\varphi$, we get
\begin{equation}
\delta=\xi^{2}\Delta_{2}+\xi^{4}\Delta_{4}+\mathcal{O}\left(\xi^{6}\right)
\qquad\text{and}\qquad
A=1+\xi^{2}\Lambda_{2}+\xi^{4}\Lambda_{4}+\mathcal{O}\left(\xi^{6}\right),
\end{equation}
with
\begin{equation}
\Delta_{2}=\int_{x}^{\pi/2}\text{d}y\left((\varphi')^{2}+(\dot{\varphi})^{2}\right)\mu\nu \qquad\text{and}\qquad \Lambda_{2}=-\nu\int_{0}^{x}\text{d}y\left((\varphi')^{2}+(\dot{\varphi})^{2}\right)\mu.
\end{equation}
Using the metric ansatz (\ref{eqn:MetricAnsatz}), we find that
\begin{equation}
L^{2}R+d(d-1)=\frac{(\cos x)^{2}e^{\delta}}{(\tan x)^{d-1}}\left(\frac{\partial C_{x}}{\partial x}-\frac{\partial C_{t}}{\partial t}-\mu'(x)e^{-\delta}(A-1)\delta'\right),
\end{equation}
with
\begin{equation}
C_{x}=e^{-\delta}\left(\mu(x)(2A\delta'-A')-\mu'(x)(A-1)-2(\tan x)^{d}A\right)
\quad\text{and}\quad
C_{t}=\mu(x)\frac{e^{\delta}\dot{A}}{A^{2}}.
\end{equation}
For the measure, one has
\begin{equation}
\int\text{d}^{d+1}\mathbf{x}\sqrt{-g}(...)=L^{d+1}V_{d-1}\int\text{d}x\int\text{d}t\,e^{-\delta}\frac{(\tan x)^{d-1}}{(\cos x)^{2}}(...).
\end{equation}
The volume of the angular part $\mathcal{S}^{d-1}$ is given by $V_{d-1}=\int_{\mathcal{S}^{d-1}}\text{d}\Omega_{d-1}=\frac{2\pi^{d/2}}{\Gamma(d/2)}$. If we expand the metric part of the action (\ref{eqn:Action}) in powers of $\xi$, we find that
\begin{align}
\mathcal{S}_{g}&=\frac{1}{16\pi G}\int\text{d}^{d+1}\mathbf{x}\sqrt{-g}\left(R+\frac{d(d-1)}{L^{2}}\right) \nonumber \\
&=\frac{L^{d-1}V_{d-1}}{2(d-1)}\int\text{d}x\int\text{d}t\left(\frac{\partial C_{x}}{\partial x}-\frac{\partial C_{t}}{\partial t}-\mu'(x)(\Lambda_{2}\Delta'_{2})\xi^{4}+\mathcal{O}\left(\xi^{6}\right)\right).
\end{align}
The Gibbons-Hawking boundary term can be written as
\begin{equation}
\mathcal{S}_{GH}=\frac{1}{8\pi G}\int_{\partial\mathcal{M}}\text{d}^{d}\mathbf{x}\sqrt{-\gamma}K=-\frac{L^{d-1}V_{d-1}}{2(d-1)}\int\text{d}t\left.\left(C_{x}-\mu'(x)(1+A)e^{-\delta}\right)\right|_{x=\frac{\pi}{2}}.
\end{equation}
The total derivative of $C_{x}$ in the bulk action will cancel with the $C_{x}$ in the boundary action. The second (divergent) term in the boundary action is removed by the counterterm
\begin{equation}
\mathcal{S}_{C}=-\frac{1}{8\pi G}\int_{\partial\mathcal{M}}\text{d}^{d}\mathbf{x}\sqrt{-\gamma}\left(\frac{d-1}{L}\right)=-\frac{L^{d-1}V_{d-1}}{(d-1)}\int\text{d}t\left.\left((d-1)\frac{(\tan x)^{d-1}}{\cos x}e^{-\delta}\sqrt{A}\right)\right|_{x=\frac{\pi}{2}}.
\end{equation}
One needs to use the fact that $A\rightarrow1$ when $x\rightarrow\pi/2$. In the end, we have 
\begin{align}
\mathcal{S}_{g+GH+C}=&-\frac{L^{d-1}V_{d-1}}{2}\int\text{d}x\int\text{d}t\left(\frac{\Lambda_{2}\Delta'_{2}}{\nu}\right)\xi^{4}+\mathcal{O}\left(\xi^{6}\right) \nonumber \\
=&-\frac{L^{d-1}V_{d-1}}{2}\int\text{d}x\int\text{d}t\left(\left((\varphi')^{2}+(\dot{\varphi})^{2}\right)\mu\nu\int_{0}^{x}\text{d}y\left((\varphi')^{2}+(\dot{\varphi})^{2}\right)\mu\right)\xi^{4}+\mathcal{O}\left(\xi^{6}\right).
\end{align}
On the other hand,
\begin{equation}
L^{2}(\partial\phi)^{2}=(\cos x)^{2}A(\phi')^{2}-\frac{(\cos x)^{2}}{A}e^{2\delta}(\dot{\phi})^{2}=(\cos x)^{2}A\left(\Phi^{2}-\Pi^{2}\right),
\end{equation}
and thus
\begin{align}
S_{\phi}=&\int\text{d}^{d+1}\mathbf{x}\sqrt{-g}\left(-\frac{1}{2}(\partial\phi)^{2}\right)=-\frac{L^{d-1}V_{d-1}}{2}\int\text{d}x\int\text{d}t\,Ae^{-\delta}\left(\Phi^{2}-\Pi^{2}\right)\mu \nonumber \\
=&-\frac{L^{d-1}V_{d-1}}{2}\int\text{d}x\int\text{d}t\left(\left((\varphi')^{2}-(\dot{\varphi})^{2}\right)\mu\,\xi^{2}+\left(\Lambda_{2}-\Delta_{2}\right)\left((\varphi')^{2}+(\dot{\varphi})^{2}\right)\mu\,\xi^{4}\right)+\mathcal{O}\left(\xi^{6}\right) \nonumber \\
=&-\frac{L^{d-1}V_{d-1}}{2}\int\text{d}x\int\text{d}t\left(\left((\varphi')^{2}-(\dot{\varphi})^{2}\right)\mu\right)\xi^{2} \nonumber \\
&-\frac{L^{d-1}V_{d-1}}{2}\int\text{d}x\int\text{d}t\left(\Lambda_{2}\left((\varphi')^{2}+(\dot{\varphi})^{2}\right)\mu+\Delta'_{2}\int_{0}^{x}\text{d}y\left((\varphi')^{2}+(\dot{\varphi})^{2}\right)\mu\right)\xi^{4}+\mathcal{O}\left(\xi^{6}\right) \nonumber \\
=&-\frac{L^{d-1}V_{d-1}}{2}\int\text{d}x\int\text{d}t\left(\left((\varphi')^{2}-(\dot{\varphi})^{2}\right)\mu\right)\xi^{2} \nonumber \\
&-\frac{L^{d-1}V_{d-1}}{2}\int\text{d}x\int\text{d}t\left(-2\left((\varphi')^{2}+(\dot{\varphi})^{2}\right)\mu\nu\int_{0}^{x}\text{d}y\left((\varphi')^{2}+(\dot{\varphi})^{2}\right)\mu\right)\xi^{4}+\mathcal{O}\left(\xi^{6}\right)
\end{align}
All in all, up to fourth order in the scalar field, we find the action
\begin{equation}
\tilde{S}=-\frac{L^{d-1}V_{d-1}}{2}\int\text{d}x\int\text{d}t\left(\left((\phi')^{2}-(\dot{\phi})^{2}\right)\mu-\left((\phi')^{2}+(\dot{\phi})^{2}\right)\mu\nu\int_{0}^{x}\text{d}y\left((\phi')^{2}+(\dot{\phi})^{2}\right)\mu\right).
\end{equation}

\section{Relating the coefficients in the renormalization and averaging procedures}
\label{Sec:CoefsRGvsAve}

Here, we present the details of the calculations that led to the result (\ref{eqn:AveHam}) for the averaged Hamiltonian $\overline{\mathcal{H}}$. Following the procedure outlined in section \ref{Sec:AveragedHamiltonian}, we find after some straightforward algebra that 
\beq
\overline{\mathcal{H}}=-\frac{\epsilon^{2}}{2}\hspace{-5mm}\sum\limits_{\begin{array}{c}\vspace{-7mm}\\\ssty ijkl\vspace{-2mm}\\\ssty\omega_{i}+\omega_{j}=\omega_{k}+\omega_{l}\end{array}}\hspace{-8mm}\Omega_{ijkl}\alpha_{i}\alpha_{j}\bar{\alpha}_{k}\bar{\alpha}_{l} -\frac{\epsilon^{2}}{2}\hspace{-5mm}\sum\limits_{\begin{array}{c}\vspace{-7mm}\\\ssty ijkl\vspace{-2mm}\\\ssty\omega_{i}+\omega_{j}+\omega_{k}=\omega_{l}\end{array}}\hspace{-8mm}
\Gamma_{ijkl}\left(\bar{\alpha}_{i}\bar{\alpha}_{j}\bar{\alpha}_{k}\alpha_{l}+\alpha_{i}\alpha_{j}\alpha_{k}\bar{\alpha}_{l}\right), 
\eeq
where the coefficients $\Omega_{ijkl}$ and $\Gamma_{ijkl}$ are given explicitly in terms of the $W^{(a,b)}_{ijkl}$ coefficients (\ref{eqn:WCoeffs}) as
\begin{align}
\Omega_{ijkl}&=6W^{(1,1)}_{(ijkl)}-\omega_{i}\omega_{j}W^{(0,1)}_{ijkl}-\omega_{k}\omega_{l}W^{(0,1)}_{klij}+\omega_{i}\omega_{k}W^{(0,1)}_{ikjl}+\omega_{i}\omega_{l}W^{(0,1)}_{ilkj} \nonumber \\
&+\omega_{j}\omega_{k}W^{(0,1)}_{kjil}+\omega_{j}\omega_{l}W^{(0,1)}_{ljki}-\omega_{i}\omega_{j}W^{(1,0)}_{klij}-\omega_{k}\omega_{l}W^{(1,0)}_{ijkl}+\omega_{i}\omega_{k}W^{(1,0)}_{jlik}+\omega_{i}\omega_{l}W^{(1,0)}_{kjil} \nonumber \\
&+\omega_{j}\omega_{k}W^{(1,0)}_{ilkj}+\omega_{j}\omega_{l}W^{(1,0)}_{kilj}+6\omega_{i}\omega_{j}\omega_{k}\omega_{l}W^{(0,0)}_{(ijkl)}
\end{align}
and
\begin{align}
3\,\Gamma_{ijkl}&=12W^{(1,1)}_{(ijkl)}-2\omega_{i}\omega_{j}W^{(0,1)}_{ijkl}-2\omega_{i}\omega_{k}W^{(0,1)}_{ikjl}-2\omega_{j}\omega_{k}W^{(0,1)}_{jkil}+2\omega_{k}\omega_{l}W^{(0,1)}_{klij} \nonumber \\
&+2\omega_{j}\omega_{l}W^{(0,1)}_{jlik}+2\omega_{i}\omega_{l}W^{(0,1)}_{iljk}-2\omega_{i}\omega_{j}W^{(1,0)}_{klij}-2\omega_{i}\omega_{k}W^{(1,0)}_{jlik}-2\omega_{k}\omega_{j}W^{(1,0)}_{iljk} \nonumber \\
&+2\omega_{k}\omega_{l}W^{(1,0)}_{ijkl}+2\omega_{j}\omega_{l}W^{(1,0)}_{ikjl}+2\omega_{i}\omega_{l}W^{(1,0)}_{jkil}-12\omega_{i}\omega_{j}\omega_{k}\omega_{l}W^{(0,0)}_{(ijkl)}.
\end{align}
The $\Omega_{ijkl}$ coefficients satisfy the symmetry relations $\Omega_{ijkl}=\Omega_{jikl}$, $\Omega_{ijkl}=\Omega_{ijlk}$ and $\Omega_{ijkl}=\Omega_{klij}$ and the $\Gamma_{ijkl}$ coefficients are totally symmetric in the first three indices. Using integration by parts, one can establish the relations
\begin{subequations}
\begin{align}
W_{ijkl}^{(0,1)}-\omega_{k}^{2}W_{ijkl}^{(0,0)}&=X_{kijl},
&W_{ijkl}^{(1,1)}-\omega_{k}^{2}W_{ijkl}^{(1,0)}&=Y_{iljk}, \\
\left(\omega_{k}^{2}-\omega_{l}^{2}\right)W_{ijkl}^{(0,0)}&=X_{lijk}-X_{kijl},
&\left(\omega_{k}^{2}-\omega_{l}^{2}\right)W_{ijkl}^{(1,0)}&=Y_{ikjl}-Y_{iljk}, \\
\left(\omega_{k}^{2}-\omega_{l}^{2}\right)W_{ijkl}^{(0,1)}&=\omega_{k}^{2}X_{lijk}-\omega_{l}^{2}X_{kijl},
&\left(\omega_{k}^{2}-\omega_{l}^{2}\right)W_{ijkl}^{(1,1)}&=\omega_{k}^{2}Y_{ikjl}-\omega_{l}^{2}Y_{iljk}.
\end{align}
\end{subequations}
The coefficients $X_{ijkl}$ and $Y_{ijkl}$ that appear here are defined in appendix \ref{Sec:RGCoefs}. These identities can be used to show that
\begin{subequations}
\begin{align}
\Omega_{llll}&=4T_{l}+4\omega_{l}^{2}\left(A_{ll}+\omega_{l}^{2}V_{ll}\right), \\
\Omega_{ilil}&=2R_{il}+2\omega_{l}^{2}(A_{ii}+\omega_{i}^{2}V_{ii}) &\text{if}\quad&i\neq l, \\
\Omega_{ijkl}&=4S_{ijkl} &\text{if}\quad&\{i,j\}\neq\{k,l\}\quad\text{and}\quad\omega_{i}+\omega_{j}=\omega_{k}+\omega_{l}, \\
\Gamma_{ijkl}&=8Q_{ijkl}=0 &\text{if}\quad&\omega_{i}+\omega_{j}+\omega_{k}=\omega_{l}.
\end{align}
\end{subequations}
In the end, after comparing with the expression (\ref{eqn:W}) for $W$, we deduce (\ref{eqn:AveHam}).

%%%%%%%%%%%%%%%%%%%%%%%%%%%%%%%%%%%%%%%%%%%%%%%%%%%%

\end{document}